\documentclass[12pt]{article}
\usepackage{amsmath}
\usepackage{graphicx}
\usepackage{enumerate}
\usepackage{natbib}
\usepackage{amsthm}
\usepackage{amssymb}
\usepackage{setspace}
\usepackage{verbatim}
\usepackage{algorithm2e}
\usepackage{xcolor}
\usepackage{subfig}
\usepackage{dblfloatfix}
\usepackage{rotating}
\usepackage{url}

\graphicspath{{}}
\newcommand{\blind}{0}

\begin{document}

\def\spacingset#1{\renewcommand{\baselinestretch}%
{#1}\small\normalsize} \spacingset{1}

%%%%%%%%%%%%%%%%%%%%%%%%%%%%%%%%%%%%%%%%%%%%%%%%%%%%%%%%%%%%%%%%%%%%%%%%%%%%%%

\if0\blind
{
  \title{\bf Stable Multiple Time Step Simulation/Prediction from Lagged Dynamic Network Regression Models\thanks{
    This work was supported in part by an ARO YIP Award \# W911NF-14-1-0577; corresponding Author Zack W Almquist (almquist@umn.edu).}\hspace{.2cm}}
  \author{Abhirup Mallik\\
    School of Statistics, University of Minnesota\\
    and \\
    Zack W. Almquist \\
    School of Statistics, University of Minnesota}
  \maketitle
} \fi

\if1\blind
{
  \bigskip
  \bigskip
  \bigskip
  \begin{center}
    {\LARGE\bf Title}
\end{center}
  \medskip
} \fi

\bigskip
\begin{abstract}
\noindent
Recent developments in computers and automated data collection strategies have greatly increased the interest in statistical modeling of dynamic networks. Many of the statistical models employed for inference on large-scale dynamic networks suffer from limited forward simulation/prediction ability. A major problem with many of the forward simulation procedures is the tendency for the model to become degenerate in only a few time steps, i.e., the simulation/prediction procedure results in either null graphs or complete graphs. Here, we describe an algorithm for simulating a sequence of networks generated from lagged dynamic network regression models DNR(V), a sub-family of TERGMs.  We introduce a smoothed estimator for forward prediction based on smoothing of the change statistics obtained for a dynamic network regression model. We focus on the implementation of the algorithm, providing a series of motivating examples with comparisons to dynamic network models from the literature. We find that our algorithm significantly improves multi-step prediction/simulation over standard DNR(V) forecasting. Furthermore, we show that our method performs comparably to existing more complex dynamic network analysis frameworks (SAOM and STERGMs) for small networks over short time periods, and significantly outperforms these approaches over long time time intervals and/or large networks.
\end{abstract}

\noindent%

\spacingset{1.5}
\section{Introduction}

Dynamic network analysis, prediction and simulation has a long history in statistics, computer science and the sciences (e.g. \cite{almquistpa,Almquist2014}; \cite{farmer1987adaptive}; \cite{foulds.et.al:jmlr:2011}; \cite{casteigts2011time}; \cite{goetz2009modeling}; \cite{hanneke2010}; \cite{kolar2010estimating}; \cite{krivitsky12}; \cite{leskovec2008dynamics}; \cite{snijders05,snijders96}; \cite{PhysRevE.69.065102}). Interest in dynamic systems arises from change in either the relation of interest (e.g., friendship) or the nodes (e.g., individuals). In the social sciences, dynamic network models have been used to understand important issues of disease transmission \cite[e.g., sexual contact networks][]{morris1993epidemiology}, information transmission \cite[e.g., communication during disasters][]{butts2008relational}, and other important phenomena \cite[e.g., friend formation, peer influence, etc.][]{mcfarland2014network,centola2010spread}. In statistics and computer science new methods and models have been developed for understanding panel data \cite[e.g.,][]{kolar2010estimating}, sampled data \cite[e.g.,][]{Ahmed21072009,almquistdynamic} and continuous time data \cite[][]{butts2008relational}. In the physical sciences and engineering, dynamic network models have been employed to understand server load, and other complex systems. Recently, the advent of ``Big Data" -- i.e., large-scale behavioral trace data -- have increased the interest in scalable models such the lagged logistic regression models introduced by \citet{robins2001random} and expanded by \citet{hanneke2010}, \citet{cranmer2010inferential,desmarais2010consistent,leifeld2015temporal}, \citet{almquist14,almquistdynamic}. Methods for multi-step forecasting and simulation from classic DNR models has historically been quite limited, but has ready applications in the social sciences (e.g., agent-based modeling \citep{helbing2012agent}, prediction \citep{liben2007link} and simulation based experimentation \citep{rahmandad2012reporting}) as well as applications to computer science and engineering (e.g., predicting server load \citep{prodan2009prediction}).

\begin{figure}[t]
  \centering
\subfloat[]{
  \includegraphics[width=44mm]{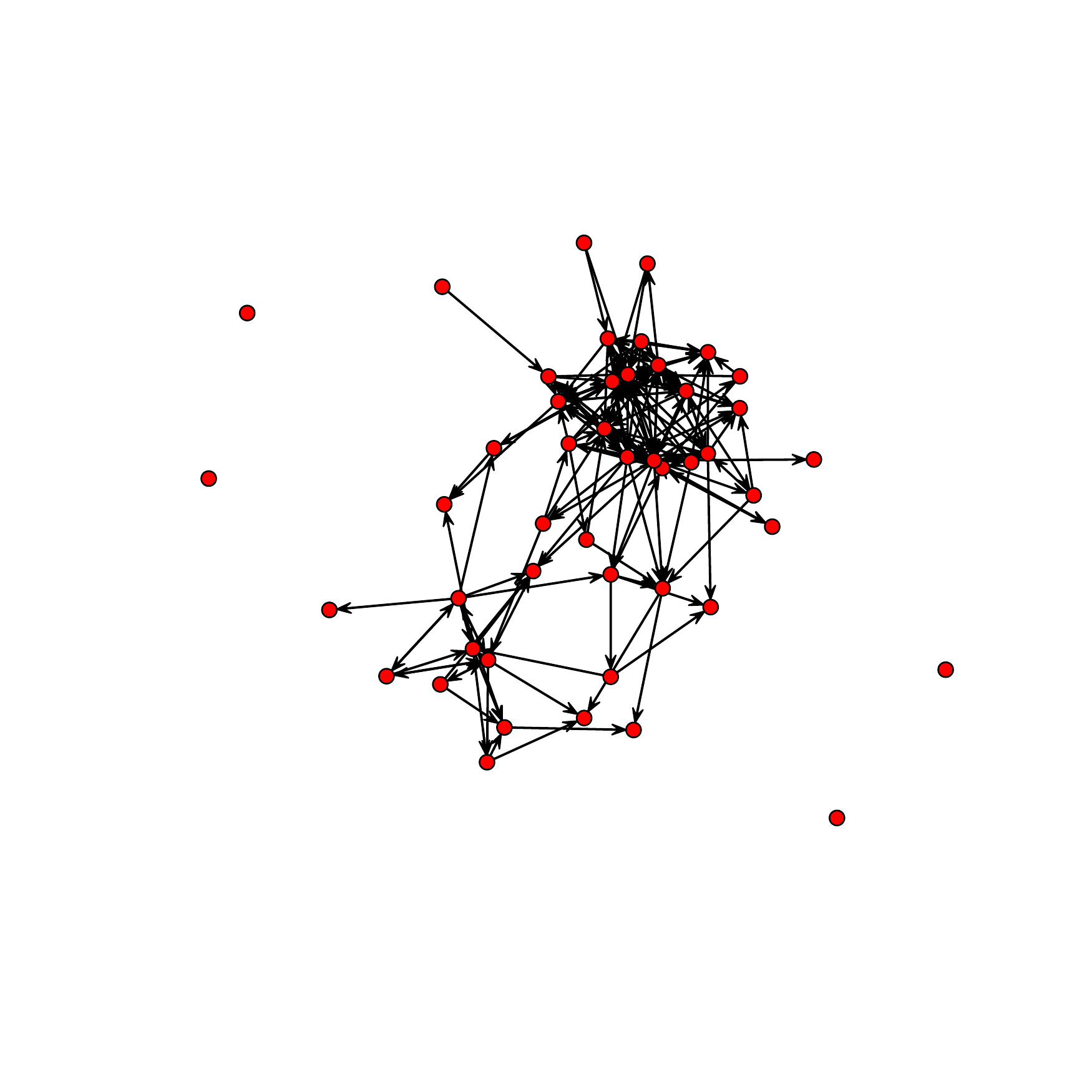}
}
\subfloat[]{
  \includegraphics[width=44mm]{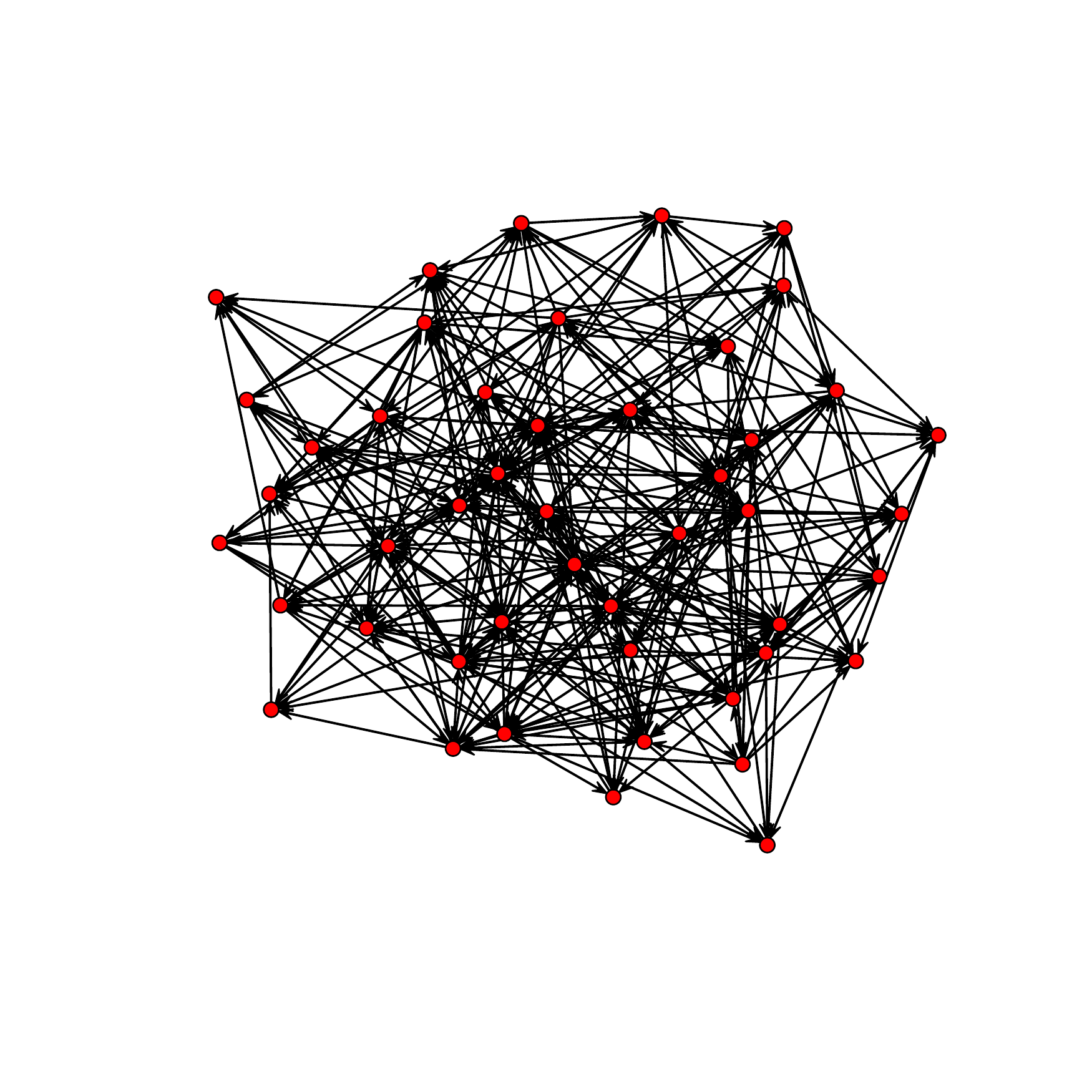}
}
\hspace{0mm}
\subfloat[]{
  \includegraphics[width=44mm]{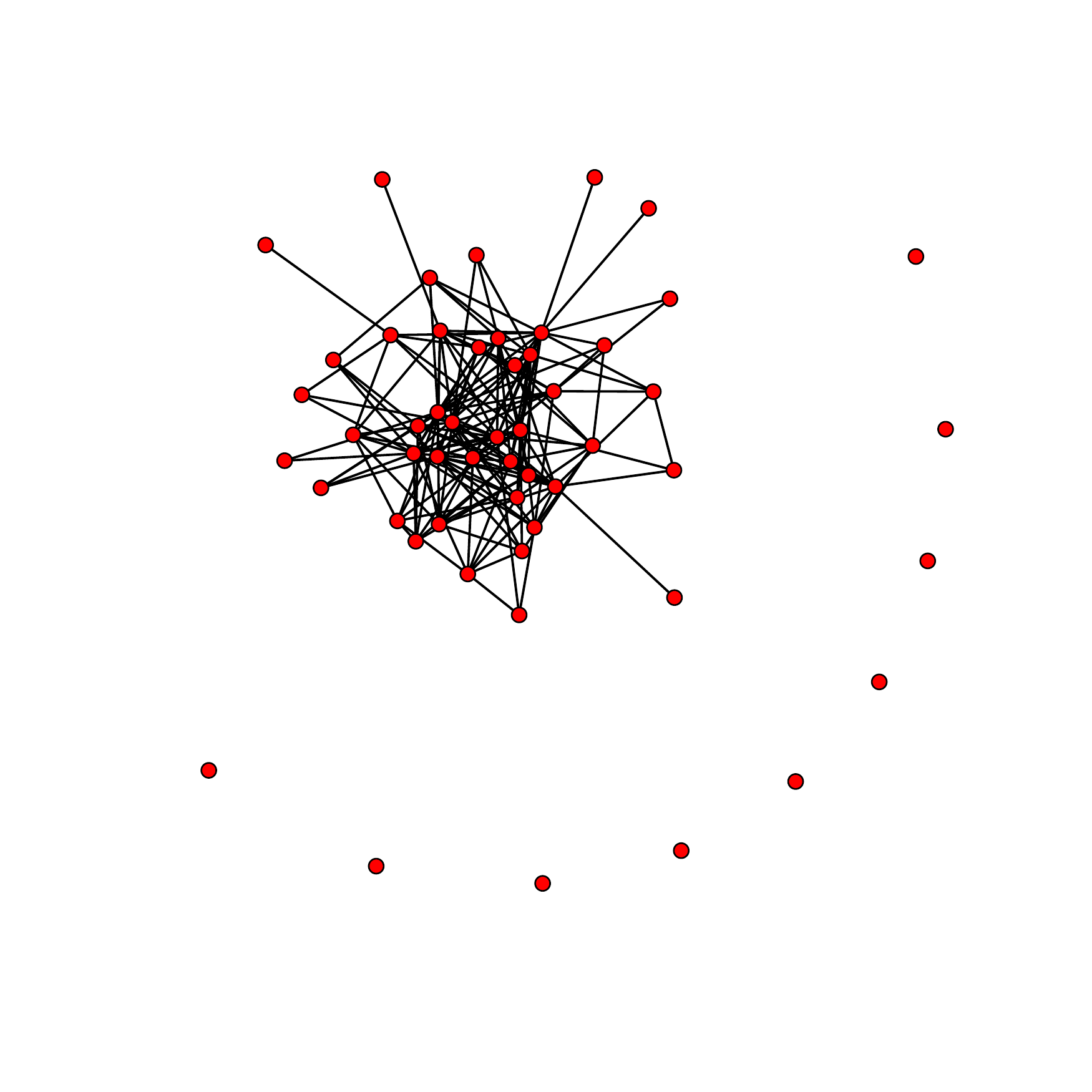}
}
\subfloat[]{
  \includegraphics[width=44mm]{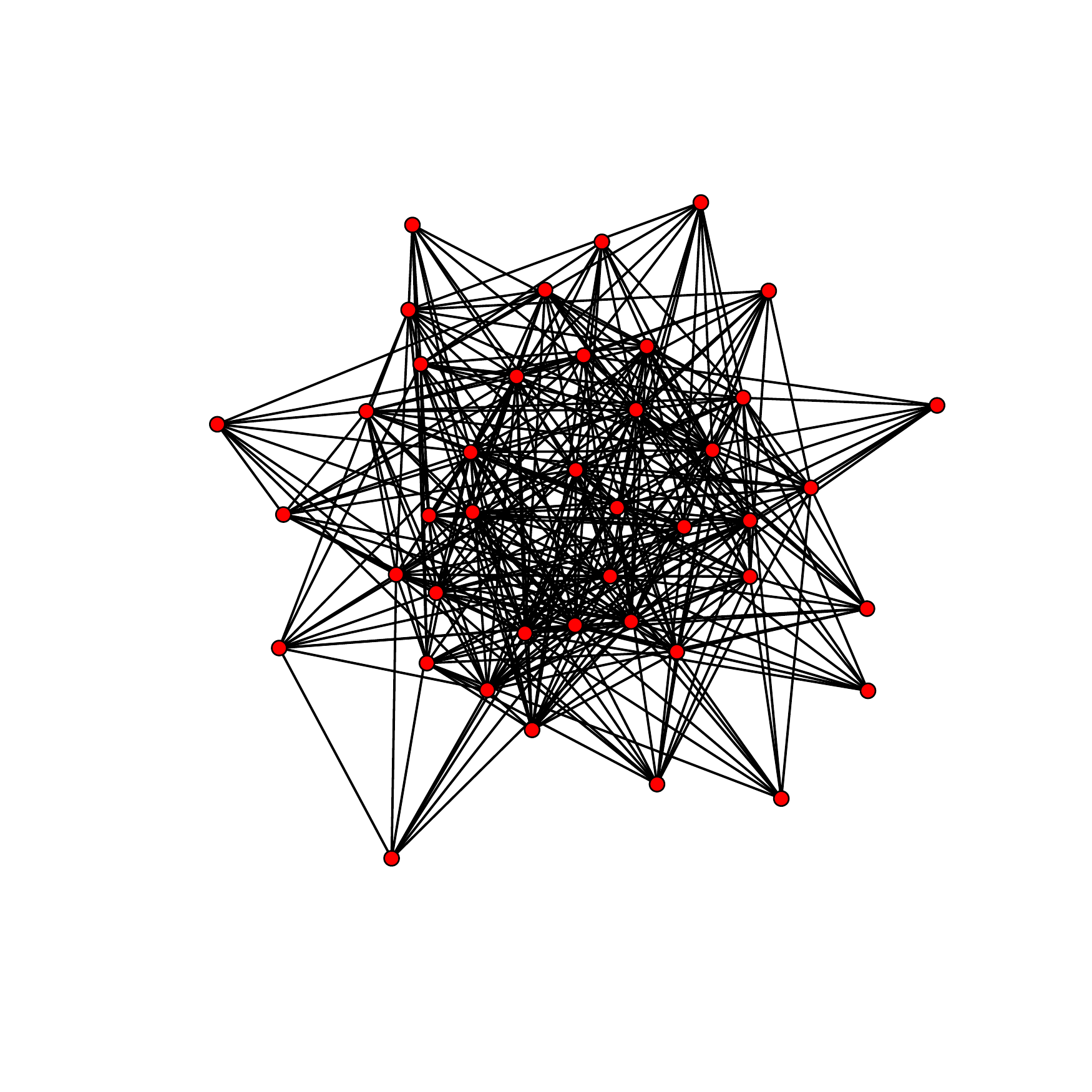}
}
  \caption{Comparison of simulated networks with or without smoothing (a) simulated network for fixed vertex case at time point 50 with smoothing, (b) simulated network at time point 50 for fixed vertex case without smoothing, (c) variable vertex case with smoothing, (d) variable vertex case without smoothing.}  \label{fig:sm-vs-ns}
\end{figure}

Lagged Dynamic Network Logistic-Regression (DNR) models provide a scalable framework for inference on large scale temporal networks collected as panel data \cite[e.g.,  network data collected hourly, daily, weekly, monthly, etc.;][]{almquistpa}. Further, DNR models readily allow for missing data \citep{almquistdynamic}  and vertex dynamics (DNR(V); e.g., change in the network via population dynamics; \citet{Almquist2014}). DNR(V) models are a subset of the \emph{Temporal Exponential-family Random Graph Models} (TERGM) \citep{hanneke2010} and are employed in computer science \citep{kolar2010estimating}, social science and the physical sciences \citep{almquistpa,desmarais2010consistent} to great effect. Further, DNR models are conceptually similar to vector autoregressive (VAR) models and depend only on the past and exogenous variables, and therefore do not require information on the current time point such as the general TERGM case which makes them an ideal framework for problems of prediction.
%Of interest to network researchers and practitioners is the ability to forecast forward dynamic networks derived from panel data (e.g., hourly, daily or weekly snapshots of a given network). Dynamic networks are often collected or saved as panel data for practical reasons such financial or time constraints (e.g., observational data collection or limited query pulls from an online API, e.g., \cite{mccormick2015using}).
% Researchers, also, aggregate continuous or near continuous time data into network snapshots; take for example Dan McFarland's study of classroom interaction \cite{mcfarland2001student}. In this work we are focused on the engaging problem of simulating or predicting \emph{multiple time steps} of network panel data. 
A common problem with the application of DNR(V) for either simulation modeling or prediction is that the model often lead to degenerate results (e.g., all networks are predicted to be complete or unconnected graphs) in only a few time steps \citep{hanneke2010}. This is a standard problem in the larger literature in statistical network models (for a review of this problem see the work of \citet{duijin07} or \citet{schweinberger2015local}). In figure \ref{fig:sm-vs-ns} we show the effect of our proposed smoothing algorithm on simulated networks for both fixed and dynamic vertex cases. Here we have simulated up to time point 50 in future for both cases. It is clear that without smoothing, the networks seem to become degenerate quite quickly. As the network prediction degrades future predictions become worse and worse; this can be readily observed in figure~\ref{fig:drift-vs}.

Here, we introduce a smoothed estimator for forward prediction, based on smoothing of the change statistics (See Section~\ref{sec:simulation_engine} for details) obtained from a dynamic network regression model. We focus on the implementation of the algorithm, providing a series of motivating examples with comparisons to dynamic network models from the literature. We find that our algorithm significantly improves multi-step prediction/simulation over standard DNR forecasting. Furthermore, we show that our method performs comparably to existing more complex dynamic network analysis frameworks (Stochastic Actor Oriented Models and Separable Temporal Exponential Random Graph Models) for small networks over short time periods, and significantly outperforms these approaches over long time time intervals and/or large networks.

In the following sections we begin by introducing the general TERG model and sub-family DNR (with and without vertex dynamics). Next, we  cover our setting for inference, and our smoothing algorithm for prediction and simulation.  Followed by a comparison of DNR prediction with our smoothing algorithm and without our smoothing algorithm. Then we compare the predictive properties of our algorithm against the two main competitors in the dynamic network literature --  (i) the Separable Temporal Exponential Random Graph Model (STERGM) \citep{krivitsky10}, and (ii) the Stochastic Actor Oriented Models (SAOM) \citep{snijders96} -- on key metrics in the computer science and social network literatures. Finally, we concluded with brief discussion of our findings.

\section{Dynamic Network Analysis}

A dynamic network is composed of entities (e.g., actors, respondents, computers, etc) and relations (e.g., friendship, communication, needle-sharing, etc.) and is typically represented as the mathematical object known as a graph $G=(E,V)$, where $E$ represents the set of relations and $V$ represents a set of vertices. This can be readily extended to handle time by adding an index $t$. In practice, we represent a graph as binary adjacency matrix, $[Y]_{n\times n}$, and temporal network as time indexed array of adjacency matrices (typically the diagonal is treated as 0 or NA in most settings). In this work we follow the notation of \citet{Almquist2014}. We begin by considering networks with a fixed number vertices, i.e., $|Y_t|=n$ for all time points $t$.  \cite[See][for a discussion of dynamic networks with vertex dynamics for comparison.]{Almquist2014} 
%The time series $G_{0}, G_{1}, \cdots, G_{t}, \cdots$, where $G_t = (V_t, E_t)$ denotes the adjacency matrix at time point $t$. This corresponds to $G_t = (V_t, E_t)$, the set of vertices and edges at time point $t$. To clarify the notation, $Y_t = [Y_{tij}]_{i, j \in V_t}$ denotes the adjacency matrix at time $t$. This notation will help us describe the system when both set of vertices and edges are evolving. 

In the following sections, we will use the notation $Y_t$ as a random variable denoting an adjacency matrix, with an instance of this random variable denoted by $y_t$. The shorthand notation $Y_a^b$ is used to denote the set of adjacency matrices $(Y_{a}, \cdots, Y_{b})$. We use $X_t$ to denote covariates associated with edges of a network $Y_t$. The function $s(.)$ will be used to denote the set of sufficient statistics for an ERGM model described next.

\subsection{Temporal Exponential-family Random Graph Models}

The framework for TERGMs is based on extending the classic Exponential Random Graph Models (ERGM) \citep{holland81a,handcock:ch:2003} to the temporal case via a VAR-type process with a  $k$th order temporal Markov assumption. This assumption is as follows, for all times $t$, $Y_t \ | \ Y_{t-1},\ldots,Y_{t-k}$ is independent of $Y_{t-k-1},\ldots$ \citep{Almquist2014}, and allows us to write the TERGM likelihood in the following form (following \citet{Almquist2014}'s notation):
\begin{align}
  \label{eq:tergm}
\Pr&(Y_t=y_t \: | \: Y^{t-1}_{t-k}=y^{t-1}_{t-k},X_t) =\\ 
&\frac{ \exp \left(\theta^T s(y_t,y^{t-1}_{t-k},X_t) \right) }{\sum_{y_{t}^{'} \in \mathcal{Y}_t} \exp \left(\theta^T s(y_t^{'},y^{t-1}_{t-k},X_t) \right)} \mathbb{I}_{\mathcal{Y}_t}(y_t), \nonumber
\end{align}
\noindent
for $y_t$ belonging to the support, $\mathcal{Y}_t$.  $s$ here is a vector of real-valued sufficient statistics with parameter vector $\theta$, and $Y^{t-1}_{t-k}=Y_{t-1},\ldots,Y_{t-k}$.  Notice that the denominator of (\ref{eq:tergm}) is intractable in the general case as it is in the static ERGM.

There have been two core methods of inference attempted for the TERG models: (i) treat this as a pooled ERGM problem  (likelihood), where the past time points are treated as covariates and estimate $\theta$ using MCMC-MLE \citep{geyer_markov_1991} or psuedo-likelihood methods \citep{strauss}, and (ii) treat the transitions between time points as a separable process where one distinguishes between tie formation and tie dissolution -- this is known as Separable TERGM or STERGM and will be discussed later in this paper.

\subsection{Dynamic Network Regression}

DNR as sub-family of TERGM makes the same VAR assumption, but also conditional independence given past information. This simplification has a number of advantages due to known issues in TERGMs of degeneracy \citep{schweinberger2015local,duijin07} and scalability \citep{Almquist2014}. Further, this model lends itself naturally to problems of network prediction as it does not rely on the current time-step for inference and updating and handles missing data \cite{almquistdynamic}. Here, again, we follow the language of  \citet{Almquist2014},
\begin{align}
   \label{eq:edge2}
\Pr&(Y_t\:|\: Y_{t-k}^{t-1}, \theta, s, X_t) =\\
 &\prod_{(i,j)\in V_t \times V_t}^n \text{Bern}\left(Y_{ijt}\left|\textrm{logit}^{-1} \left(\theta^T s(Y_{t-k}^{t-1},X_t) \right)\right.\right), \nonumber 
\end{align}
where $\text{Bern}(.)$ is understood to be the Bernoulli pmf, $\mathbb{I}$ is the indicator function, $X_t$ is a covariate set (potentially including dynamic latent variables, see supplement for discussion),  $Y^{t-1}_{t-k}=Y_{t-1},\ldots,Y_{t-k}$ is the graph structure given the vertex set from time $t-k$ to $t-1$, and $s$ is the sufficient statistics for the graph with $\theta$ being the real valued parameters of interest. Typical examples of sufficient statistics constructed from the edges of a set of networks can be found in the documentation of \citet{JSSv024i03} and many of them have been implemented in the R package \textit{ergm}.%Note that this can be readily expanded to handle vertex dynamics, see Almquist and Butts \cite{Almquist2014}. 

\subsection{Dynamic Network Regression with Vertex Dynamics}
\label{sec:variable-vertex-case}

DNR can be extended to handle vertex dynamics in a natural way through a separability condition introduced by \citet{Almquist2014}. Similar to the notation of edges, we use $V_t$ to denote the set of vertices at time point $t$ for the graph $G_t = (E_t, V_t)$. Vector of sufficient statistics for the vertices is calculated using the function $W(.)$ and the corresponding coefficients for the likelihood would be denoted by a vector $\psi$. So we will represent a graph at time $t$ using $(V_t, Y_t)$. If we take $N_v = |\cup V_t|$ to be the maximal set of nodes and $Y_a^b$ denote the adjacency matrix in a time series $Y_a, \cdots, Y_b$ then we can write $P(G_t|G_{t-k}^{t-1}, \psi, \theta, w, s, X_t) = P(V_t = v_t | G_{t-k}^{t-1}, \psi, w, X_t)   \times P(Y_t = y_t| V_t = v_t, G_{t-k}^{t-1}, \theta, s, X_{t})$. Following this logic, Almquist and Butts proposed a Dynamic Network Regression with Vertex dynamics (DNRV) as a double logistic process:
\begin{align}
  \label{eq:var-vert-lik}
  &P(G_t|G_{t-k}^{t-1}, \psi, \theta, w, s, X_t) = P(V_t = v_t | G_{t-k}^{t-1}, \psi, w, X_t)   \times P(Y_t = y_t| V_t = v_t, G_{t-k}^{t-1}, \theta, s, X_{t}) \nonumber  \\ 
  & = \frac{\exp(\psi^T w(v_t, G_{t-k}^{t-1}, X_t))}{\sum_{v' \in \mathcal{V}} \exp(\psi^T w(v'_t, G_{t-k}^{t-1}, X_t))}
  \times \frac{\exp(\theta^Ts(y_t, v_t, Y_{t-k}^{t-1}, X_t))}{\sum_{y' \in \mathcal{Y}_{v_t}}\exp(\theta^Ts(y_t, v_t, G_{t-k}^{t-1}, X_t))}  \\ 
   & = \prod_{(i)\in V_t }^{N_v} \text{Bern}\left(V_{it}\left|\textrm{logit}^{-1} \left(\psi^T w(G_{t-k}^{t-1},X_t) \right)\right.\right)  \times \prod_{(i,j)\in V_t \times V_t}^n \text{Bern}\left(Y_{ijt}\left| \textrm{logit}^{-1} \left(\theta^T s(Y_{t-k}^{t-1},X_t,V_t) \right)\right.\right)   \nonumber 
\end{align}
Here, the sufficient statistics for vertex model are denoted by $w(.)$ and for the edge model by $s(.)$. The coefficients for the vertex set is given by $\phi$ and the edge model coefficients are given by $\theta$ as in the previous model. The model is conditional on fixed lag $k$, which determines the previous state of the networks given by $Y_{t-k}^{t-1}$. The set of covariates $X_t$ can encode any exogenous covariates for the model. The multiplicative nature of this model imply that they are separable for inference. So, the parameters for the vertex model and the edge model conditional on vertex can be interpreted separately. %However, this conditionality also means that the specification of the vertex model would affect the edge model significantly.

\subsection{Brief Discussion of Model Assumptions and Parameterization}

DNR(V) generally assumes that much of the graph dependence can be captured in the past and generally follows the logic of VAR type process. DNR(V) with current time points will be the classic pseudo-likelihood estimator which has been shown to have issues in estimation of both the parameter weights and their standard errors \citep{hunter2012computational}. Recently, \citet{cranmer2010inferential} and others have employed the pseudo-likelihood estimator with a bootstrap to improve parameter and standard error estimation; however, \citet{almquistpa} demonstrated that full TERGM estimated with the bootstrap estimator with standard specifications generally does not out perform the DNR(V) model based on predictive validity checks. Currently, model specification is determined through scientific theory and predictive model assessment \cite[see for examples,][]{Almquist2014}. In this work we specify the model with common statistics chosen from the social science literature which are comparable across the different statistical models so as to allow for comparability. 

In the literature, \citet{Almquist2014} considered a set of lagged statistics for both the edge set ($s$) and vertex set ($w$). The authors focused on inertia (the lag term), lagged embeddedness (network clustering effects like triangles), popularity (lagged degree effects) and exogenous covariates such as gender in both the vertex and edge sets. Specific statistics are typified by the theory or problem at hand. For example in \citet{almquist2017shifting}, the authors considered multiple lags, cluster (embeddeness) and popularity (degree) statistics.  In this work will focus a similar set of core network metrics which also have the feature of being implementable across the different dynamic network models.

\section{Simulation/Prediction from DNR} %with a Given Parametric Structure
\label{sec:simulation_engine}

Model specification is often done through expert judgement and/or theory \citep{schwarz1978estimating} along with formal statistical methods (e.g., likelihood ratio test, BIC, etc.). If we assume a known model specification and an empirical data set we can simulate or predict from this model. In practice the parameter values are typically obtained from empirical network data, which can be estimated through either MLE \citep{Almquist2014}, Bayesian \citep{almquist14} or penalized maximum likelihood methods (for a general discussion penalized methods see \cite{tibshirani1996regression,hans2009bayesian}). Depending on the complexity of the model and the lag term, the number of coefficients to be estimated can be quite large, hence it is often a good idea to employ some feature selection methods for fitting the model. In this work we employ Lasso regression \citep{friedman_regularization_2010} to both infer the parameters and perform model selection on our \emph{training} dataset and use the algorithm discussed in this section to predict the \emph{held out} network panel data. For the current algorithm we pre-specify the length of the lag term.  We use a collection of consecutive network panels approximately of the length of the maximum lag to predict the unknown network. This collection of networks will be referred to as a \emph{window} and the algorithm shifts the window forward as we make future predictions. We have explored this space through simulation and found that while larger windows improve prediction a bit, the gains do not warrant the loss of useable data. In cases where one wants to simulate from a known generative procedure where prediction of a real-world network is not the goal one may employ a static ERGM to inform the initial time-points. The initial window is selected by the researcher. In instances when we have an input sequence of networks larger than the size of the window we calculate the change statistics of each window by shifting the window through the sequence of networks.  The ``change statistics" \citep{JSSv024i03} or ``change scores" \citep{snijders.et.al:sm:2006} underly the core estimation algorithm for general ERGM estimation.
\citet{hunter2012computational} derives the change score via the odds ratio of the conditional graph for each dyad such that $Odds_{\theta} (Y_{ij} = 1 | Y-(i,j) = y-(i,j)) = \exp\{\theta^T\Delta_{ij}s(y)\}$, where $(i,j)$ is the i,jth edge. (It is also noted that this formulation allows for a ``local" interpretation of ERGMs.) We then calculate the mean of these window of change statistics as our estimate of the (average) change statistic matrix (this results in what is effectively a moving average of the change score statistics). This matrix is used as predictor in our simulation/prediction algorithm. 
%TODO: We need a formal definition of "change statistics"

\subsection{Estimation of sufficient statistics for prediction/forecasting}
\label{sec:estim-suff-stat}

Given a set of input parameters $\theta$, we would like to be able to forecast the future networks. For this, we would be using the likelihood given in equation \ref{eq:edge2} and we need to estimate the sufficient statistics $s(Y_{t-k}^{t-1})$. The least number of networks needed to estimate this sufficient statistics is at time points $(t-1), \cdots, (t-k)$, henceforth will be referred to as time window of length $k$. Hence a simple way of estimating these would be to use the network statistics calculated based on $Y_{(t-k)}, \cdots, Y_{(t-1)}$. However, we have found this estimate is not a stable one. This is expected as network statistics calculated based on one instance of the realization of underlying probabilities is subjected to noise in that realization. As this quantity is quite essential for predicting the future states of the network, a poor estimate would result in poor quality simulations. We have demonstrated this in simulation studies later in section \ref{sec:simstudy}.

Under the assumption that the model is sufficiently explaining the state of the network, we assume the set of sufficient statistics in a window of time points to be stable. In some cases this may be a strong assumption and in other cases a weaker assumption. For example, it is typical to assume stable mean degree in static and dynamic networks \citep{butts2015flexible} which would show up as a stable effect in our model. In other cases this smoothing (given an appropriately chosen window) may thought of as an approximation to true temporal effect.\footnote{Recent work by \citet{lee2017varying} attempts to loosen the assumption of homogeneity on the parameter space and lets evolve over time. Such an alternative formulation could be very useful when networks are rapidly evolving or simply out of equilibrium.}  Under these assumptions, we propose the following estimates for the network statistics.
\begin{equation}
  \label{eq:stat-est}
  \hat{s}(Y_{t-k}^{t-1}, X_t) = \frac{1}{(t-k)} \sum_{\tau = k+1}^t s(Y_{\tau - k}^{\tau - 1}, X_{\tau})
\end{equation}
We then plug in this estimator in the likelihood in equation \ref{eq:edge2} to produce the future states of the networks. Using this smoothing window we are able to obtain future predictions/simulations from our model which have have better properties than pure DNR/DNRV.  We summarize this algorithm as follows:

%This simulation method uses the parameter estimation algorithm for calculating the change statistics and we feed the smoothed estimate of the window change statistics to the function simulate.ergm to produce an estimate of the new network. 

 % in \ref{Algo:Simulation}.
% \begin{algorithm}
% 	\SetAlgoLined
% 	\SetKwInOut{Input}{input}\SetKwInOut{Output}{output}
% 	initialization\;
% 	\Input{Initial sequence of networks $(Y_1, \cdots, Y_T)$, Model formula, parameters ($\theta)$, $K$ (maximum lag)}
% 	\Output{Sequence of networks $(Y_{T+1}, \cdots, Y_{T+l})$}
% 	$l \leftarrow $ Size of initial networks\;
% 	\For{$i < \text{number of simulation}$}{
% 	$count \leftarrow 1$ \;
% 	\While{$count < (l-k)$}{
% 		$X[[count]] \leftarrow$ Change Statistics of current window\;
% 		Current Window $\leftarrow$ Right Shift Window\;
% 		$count ++$\;
% 	}
% 	$\bar{Y} \leftarrow$ Mean(Y)\;
% 	prob $\leftarrow \text{logit} (\bar{Y}^T\theta)$ \;
% 	NewNetwork[$t+1$] $\leftarrow$ simulate.ergm(prob) \;
% 	Shift initial networks to the right \;
% 	}
% 	\caption{Simulation Engine using Smoothed Moving Window Logistic Regression}
% 	\label{Algo:Simulation}
%       \end{algorithm}
      
% \redcomment{Is is alright to use ERGM simulate functions in the algorithm? else, we might need to write everything in formula}

\begin{algorithm}
  \SetAlgoLined
  \SetKwInOut{Input}{Input}
  \SetKwInOut{Output}{Output}
  \Input{$(G_1, \cdots, G_T), \theta, K, (X_1, \cdots, X_T)$}
  \Output{$G_{T+1}, \cdots, G_{T+L}$}
  \For{step l = 1 to L}{
    Estimate network statistics $\hat{s}(Y_{t+l-k}^{t+l-1}, X_t)$ using equation \ref{eq:stat-est}\;
    Generate $Y_{t+l}$ from likelihood from equation \ref{eq:edge2}\;
    Construct network $G_{t+l} = (V_{t+l}, E_{t+l})$, using adjacency matrix $Y_{t+l}$.\;
  }
  \caption{Algorithm for simulating networks in static vertex case.}
  \label{Algo:static-vertex-sim}
\end{algorithm}

%\subsection{Estimation of Coefficients}
%\label{sec:estim-coeff}

The researcher chooses the initial sequence of networks of size $l$, and our maximum lag size is $k$, we start simulating the $(k+1)$th network from the first $l$ networks. Also, we assume that the parameters supplied are obtained from the same sequence of networks of length $l$. We then use the likelihood specified in equation~\ref{eq:edge2} to estimate the coefficients in the model. As mentioned in section~\ref{sec:simulation_engine}, selecting the generative features of a complex model is a quite hard problem \citep{tibshirani1996regression}, especially for dynamic network models, where the number of coefficients can be quite large depending on the set of sufficient statistics specified. To solve this problem, here we employ $L_1$ penalized likelihood methods for model selection.
For constructing the predictor matrices for vertex and edge models, we use a moving window method and stack the matrices of network statistics together as the window moves forward. We use the notation $(w(.))_{t \in T}$ to denote the stacking operation on the matrices for the time index $t \in T$. We then use these stacked matrices in the likelihood equation~\ref{eq:var-vert-lik} to estimate the coefficients $(\theta, \psi)$. For completeness we specify the algorithm for parameter estimation in Algorithm \ref{Algo:parameter}.

\begin{eqnarray}
  \label{eq:vertex-model-stack}
  \tilde{w}(V_t, G_{t-k}^{t-1}, X_t) &=& \big(w(V_{\tau}, G_{\tau - k}^{\tau - 1}, X_{\tau}) \big)_{\tau = k+1, V_{\tau} \in \mathcal{V}}^t \\
\label{eq:vertex-model-stack2}
    \tilde{s}(Y_{t-k}^{t-1}, X_t) &=& \big(s(Y_{\tau - k}^{\tau - 1}, X_{\tau}) \big)_{\tau = k+1}^t
\end{eqnarray}

\begin{algorithm}
  \SetAlgoLined
	\SetKwInOut{Input}{Input}\SetKwInOut{Output}{Output}
	\Input{$(G_1, \cdots, G_T)$, $X_1, \cdots, X_T$}
        \Output{$\theta, \psi$}
        Construct $\tilde{w}(V_t, G_{t-k}^{t-1}, X_t)$ using equation \ref{eq:vertex-model-stack}\;
        Match corresponding vertices\;
        Construct $\tilde{s}(Y_{t-k}^{t-1}, X_t)$ using equation \ref{eq:vertex-model-stack2}\;
        Solve for $(\theta, \psi)$ $L_1$ penalized logistic regression using the likelihood given in equation \ref{eq:var-vert-lik}\;
        \caption{Algorithm for model selection and parameter estimation for Variable vertex models.}
        \label{Algo:parameter}
\end{algorithm}
%TODO: Why does this section end abruptly?
%TODO: change the figure to include all smoothing estimators
%TODO: explain about alternative smoothing estimators

It is to be noted that the smoothing estimator proposed in equation~\eqref{eq:stat-est} is just one of the possible smoothing estimators. To justify our use of mean as a smoothing estimator we compared the drift in estimates under several alternative smoothing estimators including median, minimum and maximum values of the network statistics. We have also compared with the estimate from maximum a posteriori probability by fitting a kernel density estimator to each element of the estimated network statistics. We have called this estimate as ``Mode'' as this is implementing similar idea as definition of mode.

In Figure~\ref{fig:drift-vs} we show the plots of the parameters from the 100 iteration of the simulation engine. As we can see the parameter values decay differently based on which smoothing method was used. It is clear that using the mean as smoothing estimator produces the least drift in parameters. The means of the estimated parameters of the simulated networks are reported next to the input parameters in Table~\ref{tab:simresults}. Full details will be discussed in Section~\ref{sec:simstudy}.

%\begin{center}
%\begin{table*}[b]
\begin{sidewaystable}[b]
  \centering
  \begin{tabular}{c|ccc|ccc|ccc|ccc}
    \hline\\
    Parameter & \multicolumn{3}{c}{Network Size 10} & \multicolumn{3}{c}{Network Size 20} & \multicolumn{3}{c}{Network Size 50} & \multicolumn{3}{c}{Network Size 100}\\
 & Initial & Mean & SD &  Initial & Mean & SD &  Initial & Mean & SD &  Initial & Mean & SD \\
    \hline\\
  $DNC_{11}$ & -6.28 & -5.05 & 0.54  & -5.21 & -4.39 & 0.53  & -5.21 & -4.39 & 0.53 & -5.11 & -3.13 & 0.37 \\
  $DNC_{01}$ & -6.20 & -5.04 & 0.53 & -4.87 & -4.33 & 0.52 & -4.87 & -4.33 & 0.52 & -4.36 & -3.03 & 0.30 \\
  $DNC_{10}$ & -6.26 & -5.07 & 0.54  & -4.77 & -4.28 & 0.51 & -4.85 & -4.41 & 0.48  & -4.45 & -3.16 & 0.31 \\
  $DNC_{00}$ & -6.31 & -5.09 & 0.55  & -4.77 & -4.28 & 0.51 & -4.77 & -4.28 & 0.51  & -3.95 & -2.91 & 0.30 \\
  $TriadCensus(Y_{t-1})$ & -0.03 & -0.01 & 0.01 & 0.02 & 0.04 & 0.01 & 0.02 & 0.04 & 0.01 & 0.04 & 0.08 & 0.01 \\ %tdcs3.1
  tdcs2.1 & -0.02 & -0.01 & 0.00  & 0.02 & 0.02 & 0.01 & 0.02 & 0.02 & 0.01 & 0.04 & 0.05 & 0.01 \\
  tdcs3.2 & 0.02 & -0.00 & 0.01   & 0.07 & 0.04 & 0.01 & 0.07 & 0.04 & 0.01 & 0.08 & 0.09 & 0.01 \\
  tdcs2.2 & 0.02 & -0.00 & 0.00 & 0.04 & 0.03 & 0.01 & 0.04 & 0.03 & 0.01 & 0.04 & 0.05 & 0.01 \\
  $Y_{t-1}$ & 3.51 & 3.75 & 0.20  & 3.25 & 4.98 & 0.31 & 3.25 & 4.98 & 0.31 & 3.63 & 5.46 & 0.28 \\
  $Y_{2}$ & 6.90 & 4.28 & 0.87  & 8.13 & 5.14 & 0.43 & 8.13 & 5.14 & 0.43  & 8.19 & 5.45 & 0.39\\
\hline
  \end{tabular}
  \caption{Simulation results from various scenarios, with initial network size of $(10, 20, 50, 100)$}
  \label{tab:simresults}
\end{sidewaystable}
%\end{center} \mbox{}

\subsection{Bayesian Extension}
Following similar development as \citet{almquist14}, the likelihood equation in equation~\eqref{eq:var-vert-lik} also allows us to do Bayesian inference in the usual way. We are interested in the posterior of $(\theta, \psi)$ given $Y_1, \cdots, Y_t$. The posterior can be written as:
\begin{equation}
  \label{eq:bayesian-likelihood-1}
  \begin{split}
    P(\psi, \theta | G_1^t, s, w, X) \propto& P(\psi, \theta | s, w, X) \\
    \times& \prod_{t = 1}^t P(G_t|G_{t - k}^{t -1}, \psi, \theta, w, s, X_t)
  \end{split}
\end{equation}

For simplification, we would assume the prior on edges and vertex are independent conditional on $X$. So, this allows the following:

\begin{equation}
  \label{eq:bayesian-likelihood-2}
  \begin{split}
    P(\psi, \theta | G_1^t, s, w, X) \propto& P(\psi | w, X) \times P(\theta | S, X) \\
    \times& \prod_{t = 1}^t P(G_t|G_{t - k}^{t -1}, \psi, \theta, w, s, X_t) \\
    =& P(\psi | w, X) \frac{\exp(\psi^T w(v_t, G_{t-k}^{t-1}, X_t))}{\sum_{v' \in \mathcal{V}} \exp(\psi^T w(v'_t, G_{t-k}^{t-1}, X_t))} \\
  \times& P(\theta | S, X) \frac{\exp(\theta^Ts(y_t, v_t, Y_{t-k}^{t-1}, X_t))}{\sum_{y' \in \mathcal{Y}_{v_t}}\exp(\theta^Ts(y_t, v_t, G_{t-k}^{t-1}, X_t))}
  \end{split}
\end{equation}

So, the decomposition in equation~\eqref{eq:bayesian-likelihood-2} allows us to specify the priors of the vertex and the edge model separately and use the joint likelihood from equation \eqref{eq:var-vert-lik} for calculating the posterior.

\begin{figure}[t]
  \centering
\subfloat[]{
  \includegraphics[width=42mm]{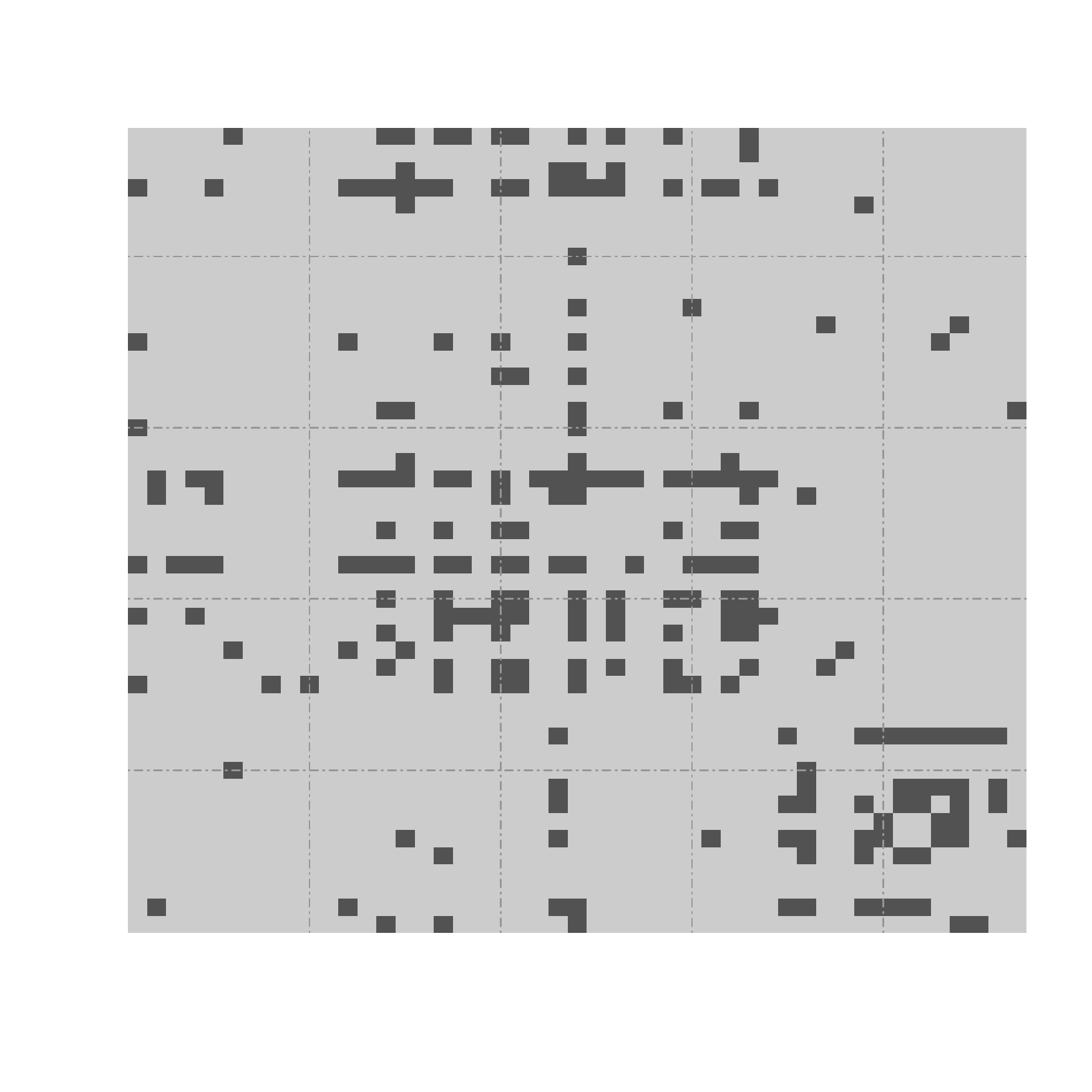}
}
\subfloat[]{
  \includegraphics[width=42mm]{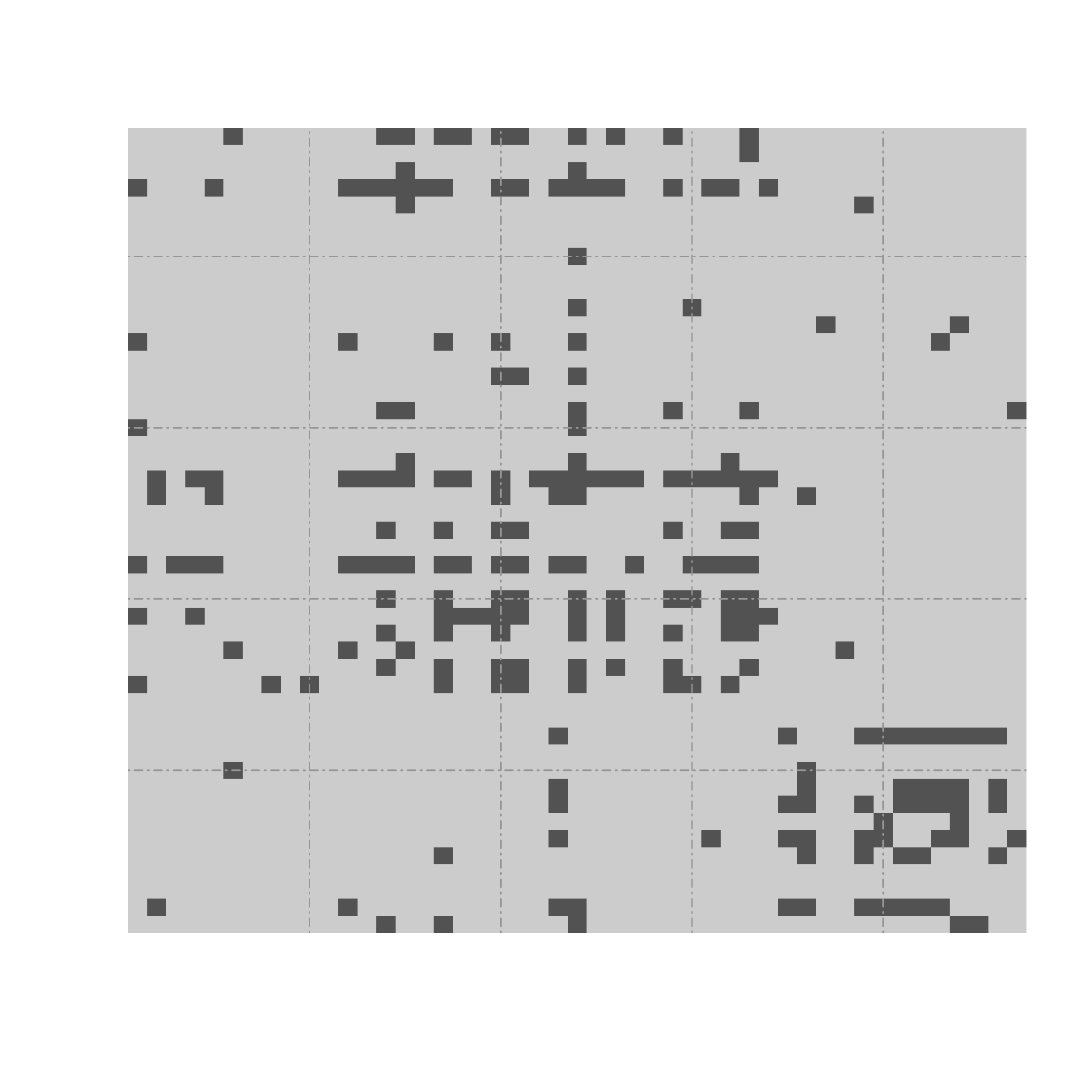}
}
\hspace{0mm}
\subfloat[]{
  \includegraphics[width=42mm]{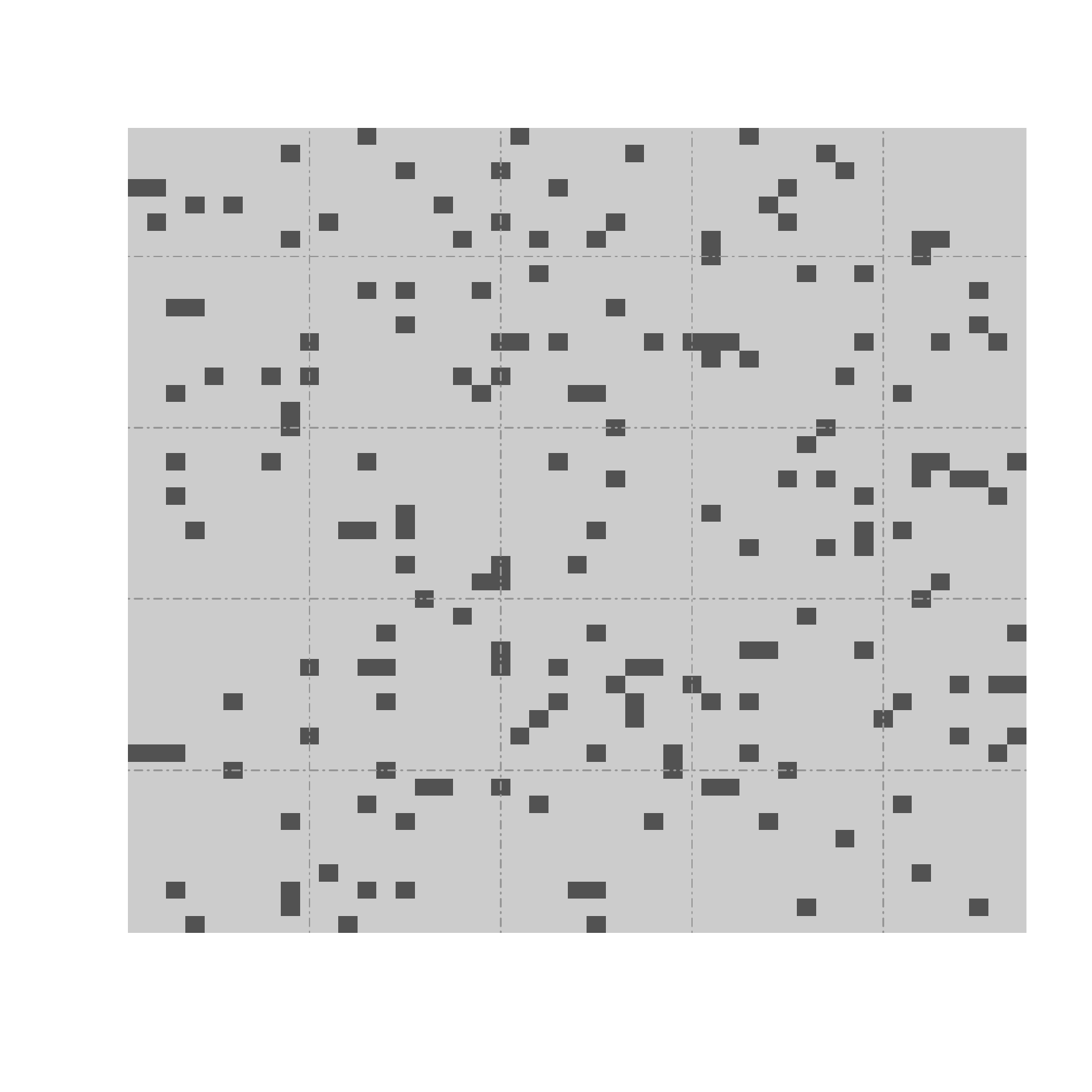}
}
\subfloat[]{
  \includegraphics[width=42mm]{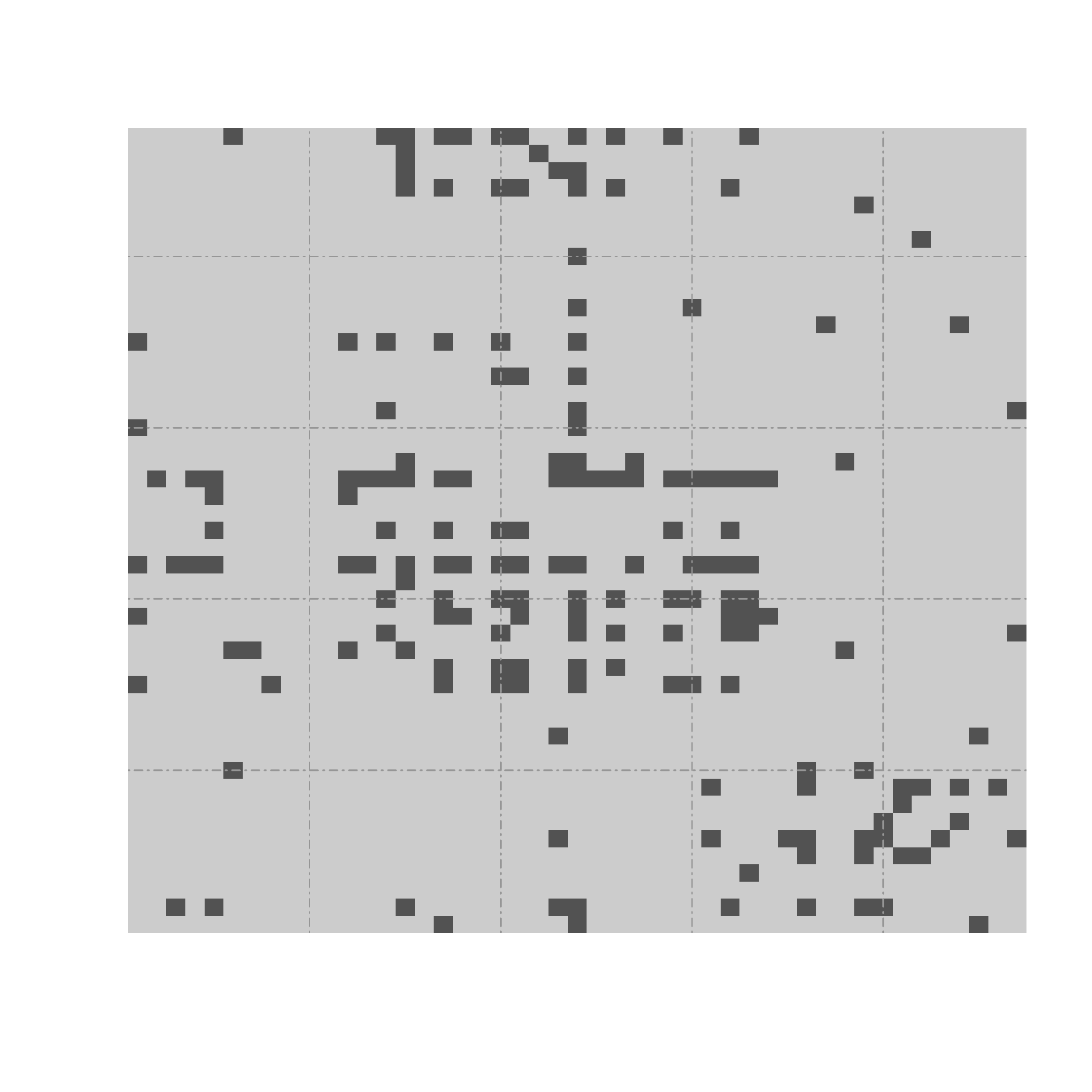}
}
  \caption{Bayesian prediction under model mis specification. (a) shows the network at time point 50, (b) shows at 51, (c) shows prediction with flat prior, (d) shows prediction with data dependent prior.}  \label{fig:bayesfig}
\end{figure}

Besides making posterior inference possible, by specifying a prior also helps us to better estimate the parameters where we have some information about them. In Figure~\ref{fig:bayesfig}, we compare the predicted networks with two kinds of priors. In this example, we have used a intercept only model with no lag terms and no other graph statistics. It can be logically argued that this is not a well suited model for this data set. We have used first $50$ time points of the blog data for training the model. We show the next step prediction of the algorithm with a flat prior, which clearly shows that the predicted network is not close to the true network; however, the use of a prior computed from the previous data points vastly improves the results. 

%%TODO:The part below should be commented
% We employ a prior on the Bernoulli parameter, which can be informative of the initial configuration of the network. Any other knowledge about the edge formation probabilities can also be encoded in the prior. As the network edges are Bernoulli, we choose the conjugate prior Beta distribution. We explain the hierarchy for $(i,j)$th edge $Y_{ij}$ here.%TODO: Clean up this equation and explain why Bayesian? Are there references for similar approach by others that we can use?

%  \begin{equation}
%    \label{eq:4}
%    \begin{split}
%      [(Y_{L+t})_{ij}|(Y_L, \cdots, Y_{L+t-1})] &\sim \text{Bern}(\theta_{ij}) \\
%      \theta_{ij} &\sim \text{Beta}(\alpha_{ij}, \beta_{ij})\\
%      (\theta_{ij}|(Y_{L+t}, \cdots, Y_{L})) &\sim \text{Beta}(\theta_{ij}+\alpha_{ij}, \beta_{ij} - Y_{ij} + 1)
%    \end{split}
%  \end{equation}

% This allows us to perform posterior inference on the simulated networks. For example the posterior mean and variance can be used to construct posterior prediction intervals. We can also use the posterior mean to include the effect of the prior in our simulated networks.
% %\redcomment{Zack: Is there a reference to penalized glm in the context of glm, that might help us explain our model selection section.}

\section{Simulation in Dynamic Vertex Case}
\label{sec:simul-vari-vert}

In the case of dynamic (variable) vertices, we are using estimated vertex regression parameters from the observed networks. The number of parameters are fixed by the model and maximum considered lag. As in the parameter estimation case, the missing vertex statistics corresponding to the vertices were imputed with zeros. The matrices of vertex statistics will be used to construct the predictors for the vertices. We use a window based average as in fixed edge case to produce smoothed estimate of the vertex statistics. The coefficients estimated from the parameter estimation process is used to produce the predictor vector for the vertices. At each time point, the vertices are simulated as a Bernoulli trial with the corresponding probabilities of presence for each vertex.

For simulating the edges conditioned on the vertices we need to first simulate the vertex set then edge set to obtain the matrix of change statistics used for parameter estimation. As, we did not use all possible edges in the regression for parameter estimation, we would not have change statistics corresponding to some edges with possibly missing vertices in some time points in the training set. So, we impute the missing edges with zero and use  mean smoothing to construct a stable estimator of the change statistics corresponding to all possible edges. They are used to get the edge probabilities using the estimated parameters. The edges are simulated from a Bernoulli trial with the edge probabilities. We keep any attributes associated with the vertices to be present in the simulated networks as well. It is assumed that the vertex attributes are not evolving with time. The set of generated edges are conditioned on the simulated vertex set. This method of generating edges conditioned on the vertices worked assuming the pattern of absence of the vertices are relatively uniform across all the time points.

For simulation in the dynamic vertex case, we need stable estimates of the network statistics for the same reasons as mentioned in section \ref{sec:estim-suff-stat}. As we use the likelihood from equation \ref{eq:var-vert-lik}, we need to estimate the vertex statistics $w(.)$ and edge statistics $s(.)$. We follow the same strategy as the static vertex case to get a stable estimate of the vertex statistics. We propose the following estimate:

\begin{equation}
  \label{eq:vert-stable-est}
  \hat{w}(V_t, G_{t-k}^{t-1}, X_t) = \frac{1}{(t-k)}\sum_{\tau = k+1, v_{\tau} \in \mathcal{V}}^{t} w(V_{\tau}, G_{\tau-k}^{\tau - 1}, X_{\tau}).
\end{equation}

Here, we use the moving average on the vertex statistics to construct the estimate of statistic at time $t$. The networks in previous time points will be of different order, hence use the set of all vertices $\mathcal{V} = \cup_{t = 1}^{T} V_t$ as the reference set of vertices. The absent vertex statistics are replaced by zero to compute the moving average equation \ref{eq:vert-stable-est}. For a stable estimate of the edge statistics $S(.)$, we use the same estimate in static vertex case in equation \ref{eq:stat-est}. Of course to calculate that estimate, we need to complete the previous networks with the bigger vertex set $\mathcal{V}$. 

The algorithm summarizing both the case for vertex and edge simulation is presented.

\begin{algorithm}
  \SetAlgoLined
  \SetKwInOut{Input}{Input}
  \SetKwInOut{Output}{Output}
  \Input{$(G_1, \cdots, G_T), \theta, \psi, K, (X_1, \cdots, X_T)$}
  \Output{$G_{T+1}, \cdots, G_{T+L}$}
  \For{step l = 1 to L}{
    Estimate vertex statistics $\hat{w}(V_t, G_{t-k}^{t-1}, X_t)$ using equation \ref{eq:vert-stable-est}\;
    Generate vertex set $V_{t+l}$ using likelihood equation \ref{eq:var-vert-lik} \;
    Estimate network statistics $\hat{s}(Y_{t+l-k}^{t+l-1}, X_t)$ using equation \ref{eq:stat-est}\;
    Generate $Y_{t+l}$ from edge likelihood conditional on $V_{t+l}$ \;
    Construct network $G_{t+l} = (V_{t+l}, E_{t+l})$, using adjacency matrix $Y_{t+l}$.\;
  }
  \caption{Algorithm for simulating networks in dynamic vertex case.}
  \label{Algo:dynamic-vertex-sim}
\end{algorithm}

%\redcomment{Zack: Before we move to the implementation and simulation part, we probably need a para explaining the novelty in our idea here.}
\section{Software Implementation}
%In practice, we implement this using the ERGM library written in R. ERGM can calculate various graph statistics from a network, using a formula interface. We require the user providing the formula that specifies the parameters in the model. The lag structure is supplied using a matrix of dimension $k \times p$, where $k$ denotes the maximum lag and $p$ is the number of model terms in the formula. For example given a formula with four terms of triadcencus and one term for curvature parameter, and maximum lag 3, we should see a matrix of size $3 \times 5$ of binary entries as an input for the lag matrix \ref{tab:LagMatrix}. Users can also input a vector of intercept terms and extraneous variables. The Lag matrix is used to specify a lag structure that the user finds most appropriate for that model and this allows for better flexibility. In the parameter estimation step we do perform model selection using $L_1$ regularized likelihood. This imposes sparsity in the lag matrix and in the vector of lag terms. For the simulation engine, we also provide time series of the estimated parameters from the generated networks for ease of diagnostics. In the Bayesian case, the user would have to provide the parameter matrices for the prior beta distribution. 

We implemented all software in {\tt R} \citep{rlang}. We use combination of custom code and software from the {\tt statnet} \citep{handcock:statnet} software-suite for computing the change score statistics and the underlying simulation engine. For the estimation and model selection we employed {\tt glmnet} \citep{friedman_regularization_2010} function, and wrote custom code posterior inference on the simulated networks. The associated R package for computation is available through Github at \url{https://github.com/SSDALab/dnr} or via CRAN \citep{dnrCRAN}. All network measures compared in Table~\ref{tab:comp} were computed using the {\tt igraph} package \citep{rigraph}.

\section{Simulation Study}
\label{sec:simstudy}

To demonstrate the usefulness of our smoothing algorithm we focus on two key features: (i) we compare the stability of our algorithm with traditional methods, and (ii) how our algorithm functions in comparison to the two main temporal network competitors in the literature. To do this we employ three classic data sets in the social network literature (described in the next subsection). To compare the forecasting ability of our algorithm with two existing methods in the literature, we take the standard machine learning approach of splitting the data in half for the training set and the holdout set. We then run all comparing methods to forecast the rest of the series. It is to be noted that for certain data sets, predicting far ahead in time can be challenging as the predictions usually converge to a limiting case. We experienced this in all the algorithms in our study, and the problem is largely one of model selection, i.e., none of the models perform well when the feature set is poorly chosen.

The next several subsections lay out the real world data under consideration. The model features we consider  for prediction within each dynamic network model, all focus on a limited set of features which are largely comparable across model types. Further, we restrict ourselves to a rather limited feature set to keep the discussion relatively parsimonious and understandable. Model prediction in all cases can be improved by considering a larger range of parameters -- however many of these models have limited scalability and cannot perform either estimation or simulation on large number of features or time points. For this reason and general call to parsimony we focus on only a small set of lagged network statistics.

\subsection{Description of the datasets}
\label{sec:description-datasets}

We focus on three publicly available data sets for our comparison study. Each is discussed below.

\subsubsection{Blog Citation Network}

The \emph{Blog Citation Network} is a temporal inter- and intra-group blog citation network collected by \citet{butts09b} and analyzed with DNR in \citet{almquistpa}. This dynamic network consists of relations (hyperlinks) between all blogs credentialed by the U.S. Democratic National Committee (DNC) or Republican National Committee (RNC) for their respective 2004 conventions. Each of these conventions are paramount for selecting their individual presidential nomination for president of the U.S. The set of actors consists of 47 nodes with 34 DNC and 14 RNC credentialed blogs and 1 credentialed in both. This data set consists of 484 time points covering  7/22/04 (shortly before the DNC convention) to 11/19/04 (shortly after the Presidential election). The data was collected in 6 hour increments consisting of the URLs linking the main page of one blot to any page within another blog. Here, we will consider various subsamples of the data We refer to this data set as the \emph{blog data}.

\subsubsection{Davis's Cocktail Party Data}

The \emph{Davis Cocktail Party Data} set is a classic social network originally collected and analyzed by \citet{davis2009deep}. The data set covers social interaction between 18 women over a period of nine months in  $14$ informal events over the aforementioned period. The data records which women met for which events. For our purpose we have collapsed the data into monthly levels, having one network for each month. In our tests, we have used $6$ months for training the models and used $3$ remaining months for prediction and comparison. We refer to this data set as the \emph{cocktail party}.

\subsubsection{Hospital Data}
The \emph{Hospital Data} is a relatively new dynamic social network originally collected and analyzed by \citet{vanhems_correction:_2013}. It contains the network of contacts between the patients, and Health Care Workers (HCW) in a hospital ward in Lyon, France. The vertices are labeled with their role within the hospital, i.e., Nurse, Patient, Doctor or Staff. The data was collected  12/6/2010 at 1:00 pm to12/10/2010 at 2:00 pm at 20 second intervals via RFID chips. We collapsed the time axis into hourly level to reduce the resolution of the data for our use. We have used first 50 time points of data for training and used next 20 time points for testing. We refer to this data set as the \emph{hospital data}.

\subsubsection{Beach Data}
\label{sec:beach-data}

The \emph{Beach Data} is a classic network data set, involving a dynamically changing network of interpersonal communications among the visitors of a Beach in Southern California. This data is observed over a one-month period producing 31 observations. There were 95 members being observed and in an average 15 people were present in one day with the maximum presence being 37. The proportion of edges in an average network were about 30\%. This data set is considered in detail by \citet{Almquist2014} describing their model of dynamic logistic regression with vertex dynamics (DNRV). We will refer to this data set as \emph{beach data}.

\subsection{Alternative Models}

While there are a number of potential formulations for dynamic network models for both estimation and simulation, we focus on the two which are in common use and implemented in software packages, both of which are have been implemented in {\tt R}, available to larger social network field. First, we consider STERGM and SAOM, and note that both these models rely on similar underlying framework to the DNR model and can be constructed within a similar parameter space for ready comparison. Further, we want to point out that both STERGM and SAOM rely on the current time points as well as the past time points which means they will have more information for both inference and prediction than the DNR model. 

\subsubsection{Separable TERGM}

The Separable TERGM was introduced by \citet{krivitsky10}, and has been implemented in {\tt R} as part of the {\tt statnet} suite of software. STERGM is based around the positing of two models for dynamic networks: one for tie formation and a second for tie dissolution. This is done through composing a a ERGM style formulation for both the formation and dissolution process. This formulation further relies on assuming that a researcher has observed two components, (i)  a cross-sectional network, and (ii) a mean relational duration. This model is fit via MCMC-MLE methods \cite[see,][for details]{tergmCRAN}. 

\subsubsection{Stochastic Actor Oriented Models}

The Stochastic AOM was originally developed by \citet{snijders96}, and published in software as SIENA \citep{siena07}. It was later made available in {\tt R} through {\tt rsiena} \citep{ripley2013rsiena}. This work has been developed substantially by \citep{snijders01,snijders97,snijders.et.al:sm:2006,snijders2011multilevel,snijders05,snijders02,mercken10}. The underlying assumptions of the Stochastic AOM is that a dynamic network arises as a cross-sectional samples from a latent continuous time Markov process in which an actor's possible ties and behavior constitute the state space -- this latent process is then simulated via a Markov Chain Monte Carlo algorithm.  \cite[For computational details see][]{ripley2013rsiena}. This framework was largely developed to de-couple questions of influence versus selection (e.g., the relationship between smoking and network structure \citep{lakon2015dynamic}).

\subsection{Model Features for Simulation Study}

Because the underlying framework for obtaining the sufficient statistics for the lagged network panel data is derived from the ERGM formulation developed for {\tt statnet} \citep{handcock:statnet} we use similar model term discussions. However, our model is described by the maximum lag period and the parameters for each lagged time period (including the present time for the STERGM and Siena models). Structurally, the model can be decomposed into five components.
\begin{itemize}
\item Fixed effects: terms that are fixed across time. Examples include the intercept (edge or density), degree or sender effects. The change statistics for these terms will be denoted by $I_{\delta}$.
\item Group: The categorical predictors for each edge (these represent homophily terms and stand in for the standard edge or density term). The change statistics for these terms will be denoted by $G_{\delta}$.
\item Model terms: The model for time $t$ network. These terms are used to specify the type of model that a static ERGM (this is used for SAOM and STERGM), for a detail descriptions of the terms possible here, we refer to the {\tt statnet} documentation \citep{JSSv024i01}. The change statistics for these terms will be denoted by $s(Y_{\delta})$. Here $s(.):\mathbb{G} \mapsto \mathbb{R}^q$ is the sufficient statistics for the network in the classic ERGM model.
\item Lagged model terms: The model terms corresponding to past networks. These models are same as the current models, however their presence is controlled by a binary matrix, called lag matrix. This allows for finer control on the model specification. The lag matrix $M \in \{0,1\}^{L \times q}$, where $L$ and $q$ are the maximum lag and the number of network statistics in the model.
\item Lagged networks: The networks from previous time point up to a finite lag. These terms will be denoted by $Y_{t-j}$ for a network of lag $j$ at time $t$. The lag terms can also be selected and this is achieved by a binary lag vector of size $L$, with 1 in the elements indicating the presence of the corresponding lag term.
\end{itemize}
Combining all the terms mentioned, we can write the model as follows:
\begin{equation}
  \label{eq:1}
  \begin{split}
     logit&[P(Y_{L+t}|(Y_{L}, \cdots, Y_{L+t-1}))] =\\
    & \sum_{i = 1}^{n_I} \beta_{0i} I_{\delta i} + \sum_{i = 1}^{n_q} \beta_{1i}G_{\delta i} +  \sum_{i=1}^q \beta_{2i}s(Y_{(L+t)\delta i}) + \\
   & \sum_{i=1}^q \sum_{j=1}^L \beta_{3i}M_{ij}(Y_{(L+t-j)\delta i}) + \sum_{j=1}^L \beta_{4j}L_j(Y_{L+t-j})
  \end{split}
\end{equation}

\subsection{Simulation Design}
\label{sec:simulation}
We conducted our simulation in four scenarios. These scenarios differ by the size of the starting network. We use initial network size of $(100, 20, 15, 10)$. The length of the predicted networks used in these four cases are $100$. We replicated this experiment $(50, 100, 100, 100)$ times to produce the standard deviation of the average estimated parameters in each case. We calculate the averages using the coefficients extracted using the given model, from the first prediction, which essentially is same as the input networks. Hence the first estimated coefficient is same as the input coefficients. We then report the mean of the time series of the coefficient series. We calculate standard deviation among all these means in different replications. We have devised $13$ different models to fit while extracting the coefficients. We report the result from one of the cases here in table~\ref{tab:simresults}.  We compare all the results in the supplementary file.

\section{Comparison of Smoothing Vs No Smoothing on Drift}
\label{sec:comp-smooth-vs}

\begin{figure}[t]
  \centering
  \includegraphics[scale = 0.50]{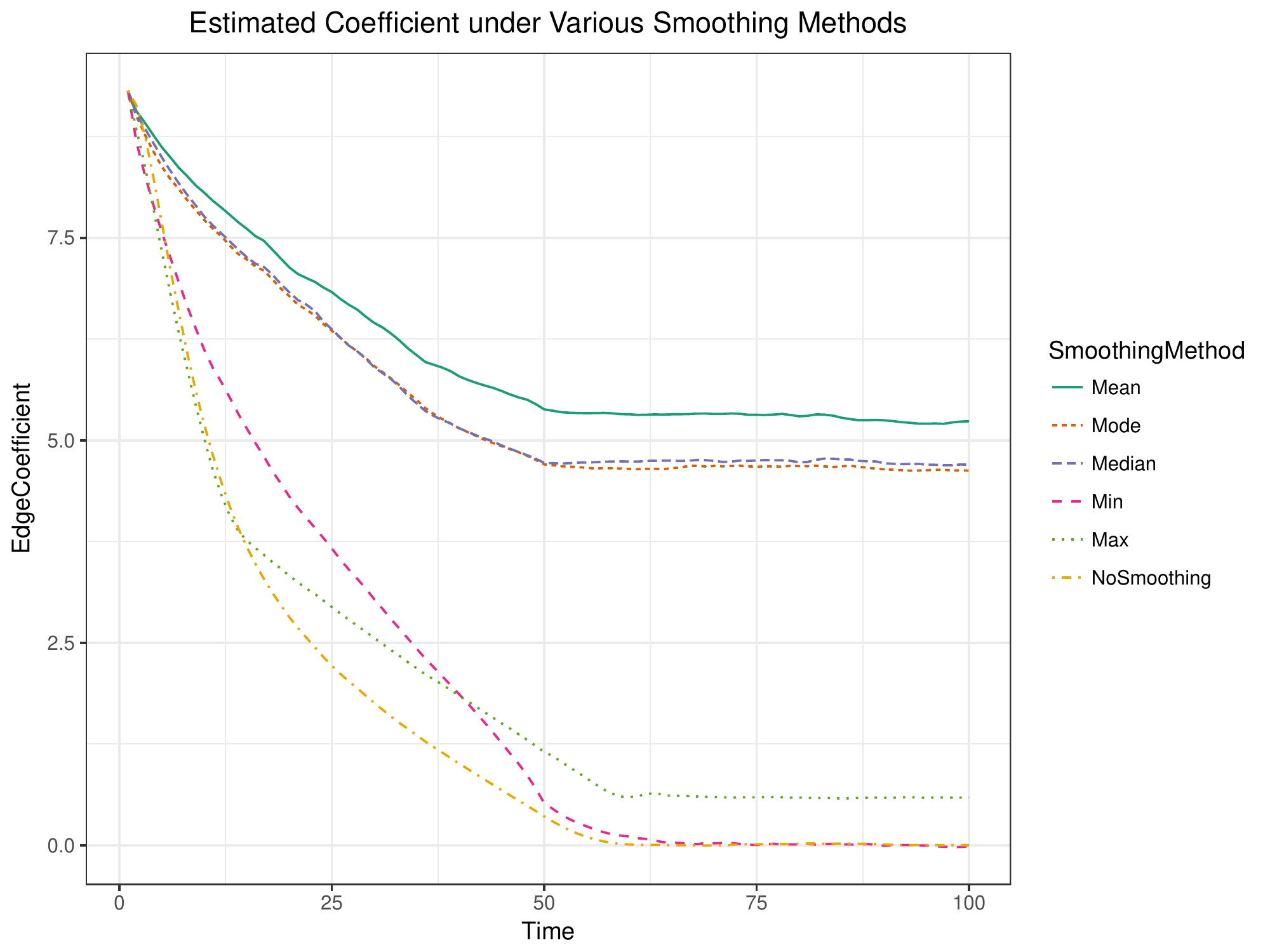}
  \caption{Comparison of simulation drift of estimates using various smoothing methods.}
  \label{fig:drift-vs}
\end{figure}

We fit a simple model of $Net \sim Edges + Lag(1)$\footnote{Using the statnet equation language.} to the blog data set with first 50 time points as the starting network. We forecast the model for next 100 time points and estimate the parameters from the forecasted series. In the version of the non-smooth prediction, we do not use the smoothed estimate of the change statistics, we only plug in the change statistics from the last step for that iteration. For the 'Mean' estimator, we use the smoothing estimator defined in equation~\eqref{eq:stat-est}. For other estimators, we use element wise operation to get the median, min and max of the estimated network statistics. In figure \ref{fig:drift-vs} we compare the time series of the estimated coefficient for the Edge parameter of the predicted networks using different smoothing estimators. The figure justifies our use of mean of the network statistics as smoothing operator on the predictor matrix to slow down the decay of the parameters, reducing the degeneracy problem of network simulation. 

\begin{figure}[t]
\centering
\subfloat[True network at T=51]{
  \includegraphics[width=40mm]{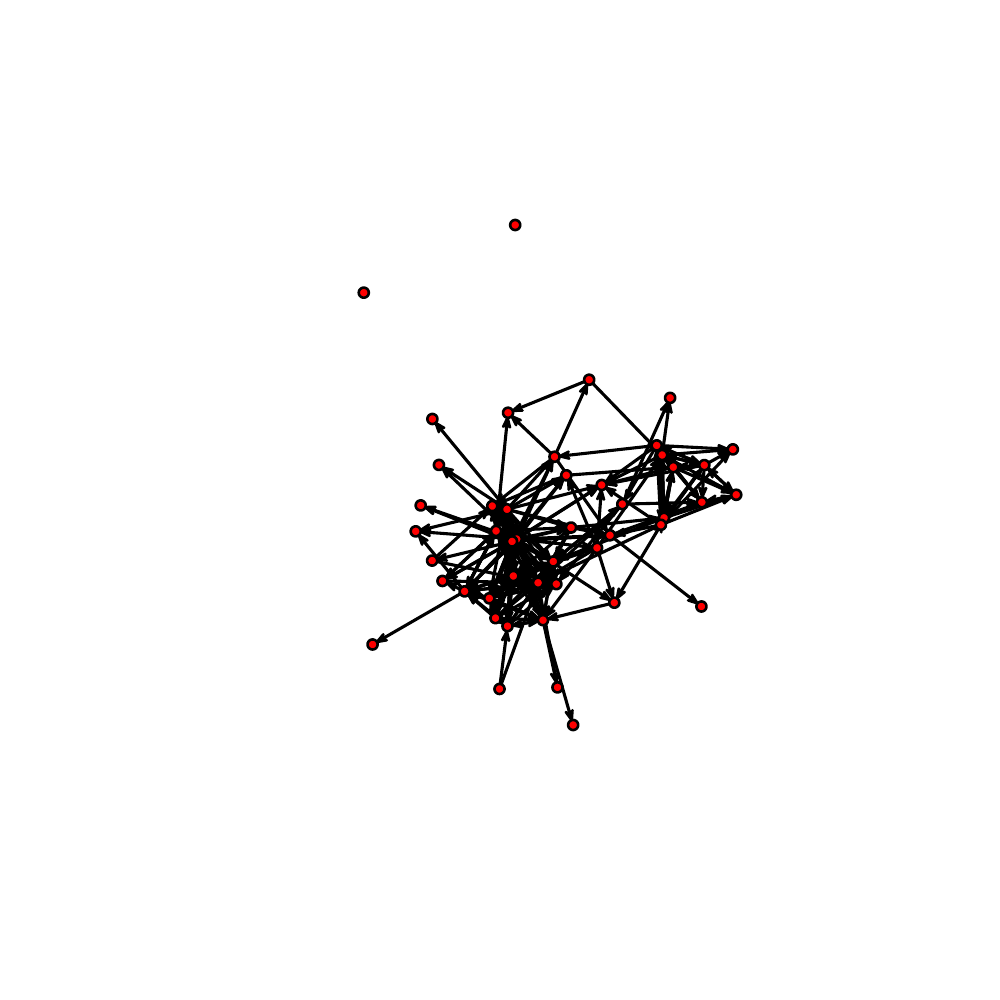}
}
\subfloat[Simulated network at T=51]{
  \includegraphics[width=45mm]{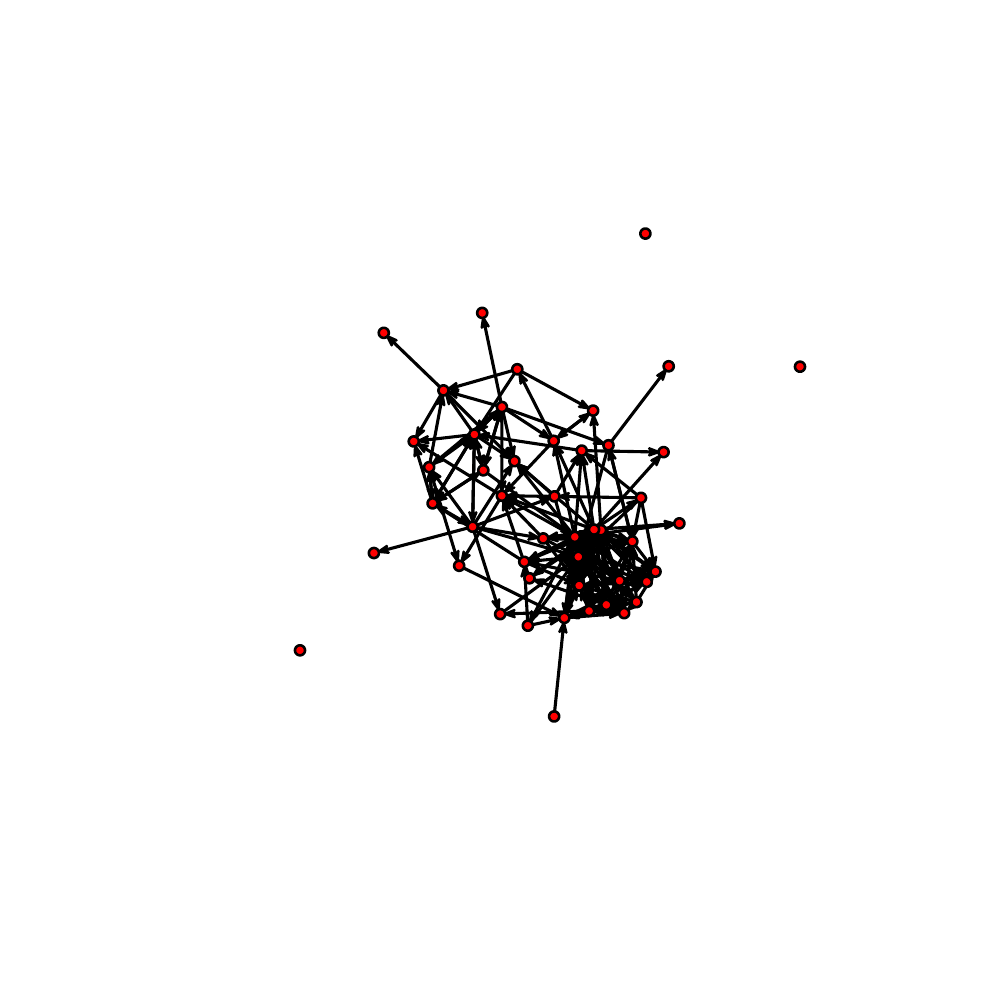}
}
\hspace{0mm}
\subfloat[True network at T=51]{
  \includegraphics[width=43mm]{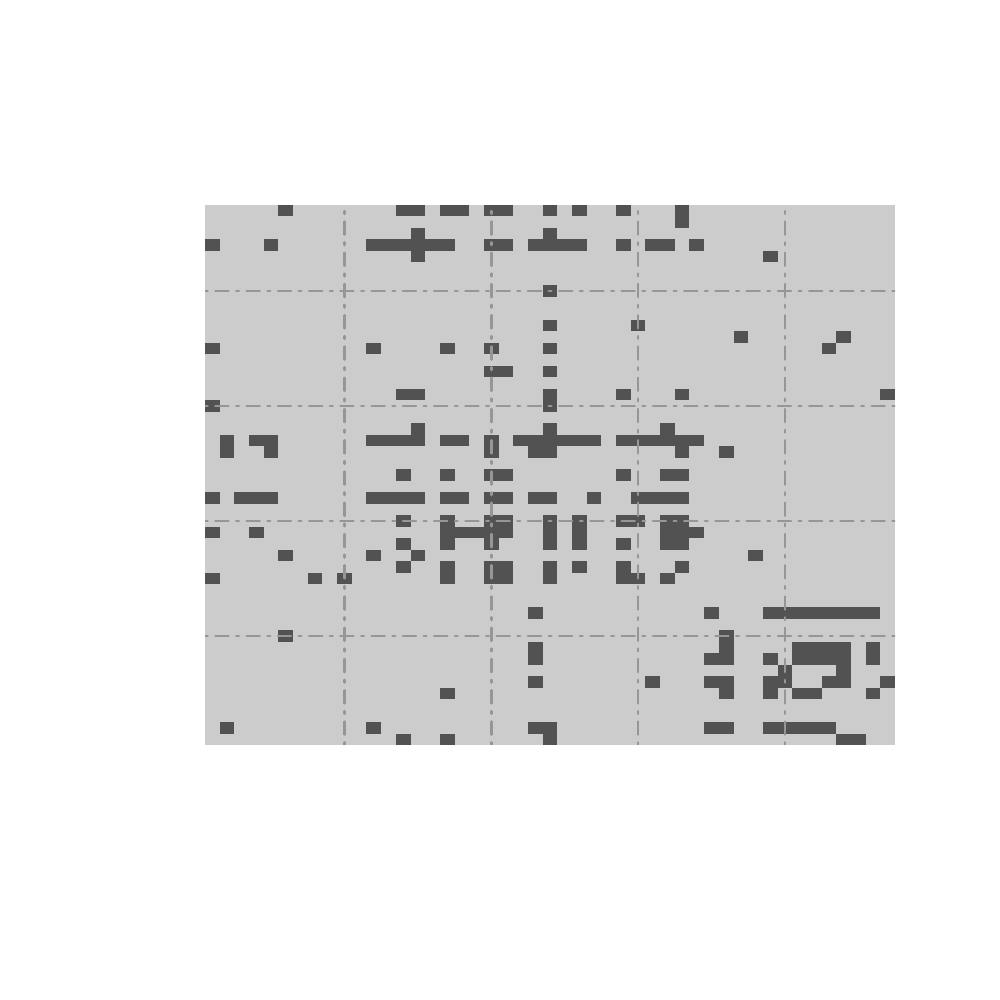}
}
\subfloat[Simulated network at T=51]{
  \includegraphics[width=43mm]{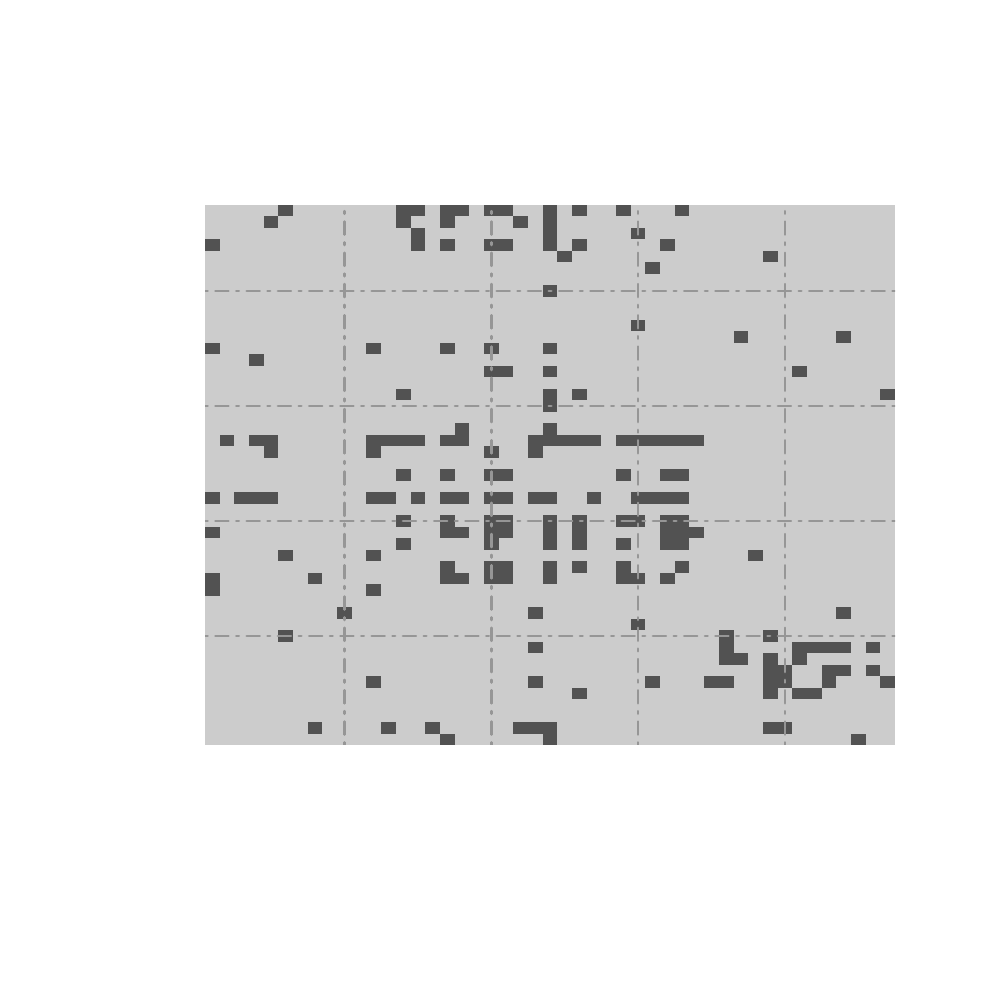}
}
\caption{Figure showing comparison of the true network and the predicted networks at first prediction time point.}
\label{fig:comp50}
\end{figure}

\section{Comparison with related methods}
\label{sec:comp-relat}

\begin{table*}[t]
\centering
\begin{tabular}{rrrrrrrr}
  \hline\\
  & \multicolumn{3}{c}{Blog Data} & \multicolumn{2}{c}{Davis Data} & \multicolumn{2}{c}{Hospital Data}\\
 & Smooth & STERGM & SAOM & Smooth & STERGM & Smooth & STERGM \\ 
  \hline
Misclassification & 0.96 & 0.89 & 0.96 & 0.95 & 0.93 & 0.99 & 0.98 \\ 
  Precision & 0.98 & 0.98 & 0.98 & 0.98 & 0.96 & 1.00 & 0.99 \\ 
  Recall & 0.77 & 0.02 & 0.73 & 0.18 & 0.09 & 1.00 & 1.00 \\ 
  $\Delta$Triangles & 62.32 & 314.76 & 61.56 & 0.00 & 0.33 & 32.95 & 32.90 \\ 
  $\Delta$ClusterCoef & 0.07 & 0.46 & 0.03 & 0 & 0.05 & 0 & 0 \\ 
  $\Delta$ExpDeg & 0.58 & 7.02 & 0.90 & 1.12 & 1.05 & 0.60 & 1.38 \\ 
   \hline
\end{tabular}
\caption{Comparison of Network simulation algorithms for three data sets. NAs represent network statistics which where not calculable under a given model. Smooth represents our smoothing algorithm for the DNR/TERG model. $\Delta$Triangles denote the absolute difference of the number of triangles with the truth. Similarly for Cluster coefficient and Expected degree.}\label{tab:comp}
\end{table*}

%Simulation of dynamic networks, based on exponential family parametrization is a problem considered by the STATNET project. The ERGM framework does have simulation options, however it does not allow for modeling lag parameters through dynamic network inputs.

%There has been several other projects that work on dynamic networks. STERGM project models the dynamic networks as a combination of formation and dissolution process, where new edges are formed or removed by a specified process. These processes can be described using ERGM formula notation. While there is no direct link between STERGM way of parameterization and ERGM.smooth, but we tried to keep the formation and dissolution process as close to our models as possible. It is also possible to target for prespecified parameter values in STERGM, which we try to use the parameters estimated in our method to keep the comparison fair.

%Another project on modeling dynamic network is Siena, which provides the R package rsiena. Even though siena uses exponential family parameterization, their network statistics are quite different from the STATNET project. Also siena uses method of moments based estimators while ERGM type of methods (including our method) rely on likelihood based estimators.

%In this document we describe several dynamic network data sets. In each of these data sets, we use half of the length of the observed data to estimate the parameters and use three different algorithms to forecast the networks. We use the holdout networks to compare with the predictions.

To compare the quality of the predictions among the algorithms we use several metrics common in the literature. The misclassification rate and precision recall metrics are calculated from the adjacency matrix of the predicted networks. We also report the number of triangles in the graphs and the clustering coefficient using the igraph library. Degree distribution of each vertex is also of importance, so we report the expected degree distribution for each graph. 

Each data set has been split into a training and testing set. The size of the training data depends on the data set and the testing set has always been immediately following the training split. For Blog citation data set, we have used first 50 days as training data, and next 10 days as testing set. For Davis's cocktail data set, we have used first 6 months as training data to predict for next 3 months. For Hospital data, we have used first 50 time points as training set and next 20 time points as testing set. To keep the comparisons among different simulation algorithms, we needed to keep the model comparable. Hence we chose simplest possible model for each data set that has common parameters in each of the methods. The common model for all data sets used edges and triad terms as sufficient statistics.

In Table~\ref{tab:comp} we compare the results using the above metrics for three simulation algorithms. In some cases the metric was not calculable, as some of the networks were possibly degenerate or the metric was not defined in those cases. We can see that the smooth ERGM algorithm performs competitively for the Blog data with SOAM (\texttt{RSiena}). In this case we used $50$ time points to train the models, and the STERGM was the worst performer. In the other two cases, We could not train Siena as it ran into singularity issues while estimating the parameters. Therefore, we only report the results for Smooth TERGM/DNR and STERGM for those cases. The smooth algorithm is performing relatively better than STERGM, especially when the training network had a small size. In figure~\ref{fig:comp50} we compare the true network at time point 51 with the simulated network using smoothing algorithm.

The CPU time for each of the methods depended on the choice of the model used. All the computation in this article has been run in a computer with dual Intel Xeon E5-2670 processor each having 16 threads, and with 32GB of RAM. In models of comparative complexity, Siena took longest time for parameter estimation and simulations, taking almost twice as much time as DNR smooth algorithm. STERGM was slightly faster than smoothed ERGM method for parameter estimation. For simulation, STERGM was also slightly faster than DNR smooth algorithm. However, we do note that CPU time comparison would depend on the parameters chosen for each specific model and the size of the input networks. In our simulation study, we also kept running parameter estimation, so the CPU times are also affected by the estimation procedure.

\section{Performance on Beach Data}
\label{sec:perf-beach-data}

We consider the \emph{beach data} as an exemplar of the dynamic vertex case. We follow the DNRV framework for estimation, and are thus able to estimate the parameters as a joint logistic process. The set of statistics for the vertex case we included are the degree of each vertex; several measures of centrality (e.g., eigen-centrality, between centrality, information centrality and closeness centrality); we also included a clustering term, in this case the count of cycles. In our final model, we kept only the significant features in each lag level only. We considered maximum of lag 3. Another important feature for the vertex only model was the group membership of the individuals. Beach goers who were regulars at the beach, had much higher chances of showing up on subsequent days.

For the conditional edge model, the important features were the lagged networks from previous time points, edge counts and the categorical nodal attribute variable on if the person was regular or not.

For selecting variables for the vertex model, we have separately modeled them as the vertex model is on the top of our model hierarchy. The selected variables is presented in table~\ref{tab:vertex-model-coefs}. The coefficient for the lag terms are all negative indicating possibly lower turnaround for consecutive days in the beach. All the coefficients for the lag terms are significant up to lag three. Interestingly, the coefficient for the indicator of week day and weekend effect is also negative. All the coefficients for the vertex parameters are positive. For variable selection for the the edge model, we had to use conditional model. The only significant terms are the lag terms and the edge count. The signs for the lag terms in the edge model is mostly different from the vertex model.

In table~\ref{tab:beachdata} we fit the model for beach data using 30 days of data and report the results on the 31st day. We have repeated the process for 30 times to compute the standard deviations.

We are, also, interested in evaluating the performance of the models across multiple prediction horizons. So, instead of using one step prediction we also simulated a sequence of networks from the prior networks. In this simulation, we have used the training data incrementally. This means, we take the first 22 days for training, then predict day 23, and in the next iteration, we take first 23 days for training and predict day 24 and so on. So, for each training data set, we predict the rest of the observed future networks and make comparisons. For this experiment, we have used days 23 to 31 for prediction of the networks. We present the results of this comparison in table \ref{tab:beach-seq-compare}. It is apparent that for the sequential simulation case the results are better than one step prediction. Large standard deviation for this case compared to one step case is resulting from the wide variation between days. The day effect was highly significant in our vertex model and was used in the simulation model to account for the variation among the weekdays and weekends.

For testing the performance of long range prediction, we forecast the next fifty time points of the beach data using both the smooth and non smooth version of the algorithm. In the non smooth version, instead of using the average predictor matrix, we only use the final value of the predictor matrix. We then compute the network statistics from the simulated networks and compare with the observed network statistics from the beach data. We present the results in table \ref{tab:sm-vs-ns-beach}. In most metrics, the smooth version of simulations are much closer to the past metrics. We also observe that the non smooth version of the algorithm produces a much denser network, with number of edges far exceeding the past number of edges. This problem is less prominent in the smooth version of algorithm. The standard deviation of the metrics from the smooth networks are also less as this produces a stabler results.

\begin{table}[ht]

\centering
\begin{tabular}{rrrr}
  \hline
 & Smooth & NonSmooth & True \\ 
  \hline
NVertices & 40.16 & 44.40 & 34.00 \\ 
 & (4.60) & (3.99) &  \\ 
  Nedges & 70.66 & 386.82 & 79.00 \\ 
 & (20.94) & (75.13) &  \\ 
  nTriangles & 21.48 & 959.52 & 91.00 \\ 
 & (13.24) & (305.55) &  \\ 
  ClustCoef & 0.14 & 0.41 & 0.59 \\ 
 & (0.04) & (0.02) &  \\ 
  ExpDeg & 7.94 & 35.54 & 10.29 \\ 
 & (1.69) & (3.93) &  \\ 
   \hline
\end{tabular}
\caption{Comparison of network metrics between smooth and non smooth algorithm for simulating network for one step prediction (time point 31).  \label{tab:beachdata} } 
\end{table}

\begin{table}[ht]
\centering
\begin{tabular}{rrrr}
  \hline
 & Smooth & Beach & NonSmooth \\ 
  \hline
Number of Vertices & 44.86 & 15.67 & 44.88 \\ 
   & (6.08) & (7.99) & (3.73) \\ 
  Number of Edges & 93.92 & 29.13 & 396.76 \\ 
   & (31.27) & (27.35) & (72.71) \\ 
 Number of Triangles & 35.62 & 35.27 & 1017.44 \\ 
   & (25.86) & (47.29) & (319.00) \\ 
  Cluster Coefficient & 0.15 & 0.65 & 0.41 \\ 
   & (0.04) & (0.15) & (0.02) \\ 
  Expected Degree & 9.23 & 7.23 & 36.11 \\ 
   & (1.94) & (3.48) & (3.68) \\ 
   \hline
\end{tabular}
\caption{Comparison between smooth and non smooth version of dynamic vertex algorithm using average of the testing data. \label{tab:sm-vs-ns-beach}}
\end{table}
%\redcomment{Zack: Do you think this performance is alright or we need to improve the model?}

\begin{table}[ht]
 
\centering
\begin{tabular}{rrrr}
  \hline
 & Triangles & ClusterCoef & ExpectedDegree \\ 
  \hline
Simulated & 85.38 & 0.71 & 35.24 \\ 
  SD & 43.89 & 0.30 & 19.47 \\ 
  True & 161.38 & 2.57 & 29.07 \\ 
   \hline
\end{tabular}
\caption{Comparison of metrics from simulation of sequence of networks using incremental training data. \label{tab:beach-seq-compare}}
\end{table}

\begin{table}[ht]
 
\centering
\begin{tabular}{rrrrr}
  \hline
 & Estimate & Std. Error & z value & Pr($>$$|$z$|$) \\ 
  \hline
lag1 & -1.6955 & 0.3374 & -5.02 & 0.0000 \\ 
  lag2 & -3.2098 & 0.5982 & -5.37 & 0.0000 \\ 
  lag3 & -2.7903 & 0.4785 & -5.83 & 0.0000 \\ 
  Day & -0.7995 & 0.0914 & -8.75 & 0.0000 \\ 
  regularLag1 & 1.4481 & 0.3401 & 4.26 & 0.0000 \\ 
  regularLag2 & 2.2312 & 0.6027 & 3.70 & 0.0002 \\ 
  regularLag3 & 1.7980 & 0.4793 & 3.75 & 0.0002 \\ 
  EigenCentralityLag1 & 1.0703 & 0.2910 & 3.68 & 0.0002 \\ 
  ClosenessCentralityLag1. & 2.3730 & 0.7405 & 3.20 & 0.0014 \\ 
  EigenCentralityLag2. & 1.0499 & 0.3190 & 3.29 & 0.0010 \\ 
  EigenCentralityLag3. & 0.9734 & 0.3223 & 3.02 & 0.0025 \\ 
   \hline
\end{tabular}
\caption{Parameters for Vertex model for Beach data. \label{tab:vertex-model-coefs}}
\end{table}

\begin{table}[ht]
  
\centering
\begin{tabular}{rrrrr}
  \hline
 & Estimate & Std. Error & z value & Pr($>$$|$z$|$) \\ 
  \hline
Edges & -3.3624 & 0.0397 & -84.77 & 0.0000 \\ 
  Lag1 & -0.1140 & 0.2098 & -0.54 & 0.5870 \\ 
  Lag2 & 1.1424 & 0.1721 & 6.64 & 0.0000 \\ 
  Lag3 & 5.1041 & 0.1101 & 46.35 & 0.0000 \\ 
   \hline
\end{tabular}
\caption{Parameters for conditional edge model for Beach data.\label{tab:edge-model-coefs}}
\end{table}

\begin{table}[ht]
\centering
\begin{tabular}{rrrrr}
  \hline
 & Estimate & Std. Error & z value & Pr($>$$|$z$|$) \\ 
  \hline
  lag1 & -1.7012 & 0.3423 & -4.97 & 0.0000 \\ 
  lag2 & -3.1539 & 0.5976 & -5.28 & 0.0000 \\ 
  lag3 & -2.8204 & 0.4802 & -5.87 & 0.0000 \\ 
  Day & -0.8550 & 0.1005 & -8.51 & 0.0000 \\ 
  attrib1 & 1.3694 & 0.3451 & 3.97 & 0.0001 \\ 
  attrib2 & 2.1787 & 0.6028 & 3.61 & 0.0003 \\ 
  attrib3 & 1.7883 & 0.4802 & 3.72 & 0.0002 \\ 
  Vstat4Lag1. & 1.2435 & 0.3013 & 4.13 & 0.0000 \\ 
  Vstat7Lag1. & 2.4526 & 0.7467 & 3.28 & 0.0010 \\ 
  Vstat4Lag2. & 0.9250 & 0.3271 & 2.83 & 0.0047 \\ 
  Vstat4Lag3. & 1.0612 & 0.3263 & 3.25 & 0.0011 \\ 
  \hline
  \hline
  edges & -0.5421 & 0.5456 & -0.99 & 0.3205 \\ 
  edgecov.regular11 & 0.0001 & 0.0003 & 0.37 & 0.7087 \\ 
  edgecov.regular00 & -0.0006 & 0.0004 & -1.47 & 0.1418 \\ 
  edgecov.regular10 & 0.0000 & 0.0003 & 0.01 & 0.9960 \\ 
  logCurrNetSize & -0.6399 & 0.1524 & -4.20 & 0.0000 \\ 
  dayEffect & 0.5746 & 0.0658 & 8.74 & 0.0000 \\ 
  lag2 & 4.6833 & 0.0866 & 54.08 & 0.0000 \\ 
   \hline  
\end{tabular}
\caption{Parameters used for one step prediction }
\end{table}

\section{Discussion}

Here, we have introduced a novel technique for improving network simulation and prediction for DNR(V) models and finally we have compared these results against the current state of the art in statistical models for dynamic network data. In addition, We have, as far as the authors are aware, been the first to use penalized likelihood methods for model selection in DNR(V) framework in contrast with typical AIC/BIC methods employed currently in the field. Given that quality of feature selection in ERGMs and TERGMs is based on in-sample prediction of macro-level graph statistics (typically those not in the feature set), it suffices to show good predictive validity for demonstrating the usefulness of this technique for model selection. The appeal of this formulation for model selection is that it is more readily scalable as penalization performs model selection in a single model run, and thus does not require one to attempt to a full factorial design (or sub-sample) of possible model parameters which could be quite large (e.g., lag statistics $k$, graph statistics $l$ and exogenous variables $v$). As \citet{tibshirani1996regression} advised in seminal work one can apply the penalized methods to obtain parsimonious model by dropping the features that have weights approximately zero and refining the model with either standard likelihood based or Bayesian methods to obtain unbiased estimates of the parameters. 

We find that DNR(V) performs favorably in comparison to STERGM and SAOM such that when SAOM and STERGM are performing at their best DNR(V) does comparable and when they are at their worst DNR(V) performs better on average. Computation time is always a very important aspect of dynamic network modeling and DNR(V) compares well to both models, though is a bit slower to STERGM (though we suspect this is due to its weaker integration with \texttt{ergm} package in \texttt{R} and graph statistics chosen). 

This paper extends the DNRV model introduced by \citet{almquist14} by improving its ability to simulate and predict in comparison to simple DNRV originally introduced. We believe this method will set the basic bar for which future dynamic vertex models will have to clear in the area of simulation and prediction.

Lastly, this method adds to the dynamic network literature and allows for the direct simulation and prediction of dynamic networks from DNR(V) models.  Our results demonstrate that our method improves the prediction/simulation of multiple time steps from a DNR(V) process. This will allow in the future the ability to perform detailed sensitivity tests to measurement process underlying dynamic network data collection and for simulation based experiments centered around dynamic network data (e.g., collaboration  during a disaster \citep{butts2008relational} or communication patterns over time).

\bibliographystyle{chicago}
\bibliography{references}
\end{document}